\documentclass[sigconf]{acmart}
\pdfoutput=1
\usepackage{array}
\usepackage{booktabs}
\usepackage{multirow}
\usepackage{makecell}
\usepackage{graphicx}
\usepackage{colortbl}
\usepackage{pifont}
\usepackage[table]{xcolor}
\usepackage{siunitx}            
\usepackage{amsmath}            
\usepackage{fontawesome5}       
\usepackage{adjustbox}          
\usepackage{soul}
\usepackage{enumitem}

\soulregister\cite7
\soulregister\citep7
\soulregister\citet7
\soulregister\ref7
\soulregister\pageref7
\usepackage{tabularx}           
\usepackage{array}              

\newcommand{\imagewidth}{3cm}
\AtBeginDocument{%
  }

\setcopyright{acmlicensed}
\acmConference[SIGMOD'26]{the 2026 International Conference on Management of Data}{May 31 - June 05, 2026}{Bengaluru, India}
\acmDOI{10.1145/3786640}

\begin{document}

\title{DIVER: A Robust Text-to-SQL System with \textbf{D}ynamic \textbf{I}nteractive \textbf{V}alue Linking and \textbf{E}vidence \textbf{R}easoning}

\definecolor{r1}{HTML}{9999ff}
\definecolor{r2}{HTML}{aff5ff}
\definecolor{r3}{HTML}{ffed99}
\newcommand{\addrone}[1]{\sethlcolor{r1}{#1}}
\newcommand{\addrtwo}[1]{\sethlcolor{r2}{#1}}
\newcommand{\addrthr}[1]{\sethlcolor{r3}{#1}}
\newcommand{\linerone}[1]{\setulcolor{r1}{#1}}
\newcommand{\linertwo}[1]{\setulcolor{r2}{#1}}
\newcommand{\linerthr}[1]{\setulcolor{r3}{#1}}
\newcommand{\otheradd}[1]{\sethlcolor{red!20}{#1}}

\author{Yafeng Nan}
\email{nanyafeng@bupt.edu.cn}
\orcid{0009-0001-2145-4605}
\affiliation{%
  \institution{Beijing University of Posts and Telecommunications}
  \city{Beijing}
  \country{China}
}

\author{Haifeng Sun}
\email{hfsun@bupt.edu.cn}
\orcid{0000-0003-3072-7422}
\affiliation{%
  \institution{Beijing University of Posts and Telecommunications}
  \city{Beijing}
  \country{China}
}

\author{Zirui Zhuang}
\authornotemark[1]
\email{zhuangzirui@bupt.edu.cn}
\orcid{0000-0003-3345-1732}
\affiliation{%
  \institution{Beijing University of Posts and Telecommunications}
  \city{Beijing}
  \country{China}
}

\author{Qi Qi}
\email{qiqi8266@bupt.edu.cn}
\orcid{0000-0003-0829-4624}
\affiliation{%
  \institution{Beijing University of Posts and Telecommunications}
  \city{Beijing}
  \country{China}
}

\author{Guojun Chu}
\email{chuguojun@bupt.edu.cn}
\orcid{0000-0003-2352-4731}
\affiliation{%
  \institution{Beijing University of Posts and Telecommunications}
  \city{Beijing}
  \country{China}
}

\author{Jianxin Liao}
\email{jxlbupt@gmail.com}
\orcid{0000-0003-1486-0573}
\affiliation{%
  \institution{Beijing University of Posts and Telecommunications}
  \city{Beijing}
  \country{China}
}

\author{Dan Pei}
\email{peidan@tsinghua.edu.cn}
\orcid{0000-0002-5113-838X}
\affiliation{%
  \institution{Tsinghua University}
  \city{Beijing}
  \country{China}
}

\author{Jingyu Wang}
\email{wangjingyu@bupt.edu.cn}
\orcid{0000-0002-2182-2228}
\affiliation{%
  \institution{Beijing University of Posts and Telecommunications}
  \city{Beijing}
  \country{China}
}

\authornote{corresponding authors.}

\renewcommand{\shortauthors}{Yafeng Nan et al.}

\begin{abstract}

In the era of large language models, Text-to-SQL, as a natural language interface for databases, is playing an increasingly important role. State-of-the-art Text-to-SQL models have achieved impressive accuracy, but their performance critically relies on expert-written evidence. This evidence typically clarifies schema and value linking that existing models \otheradd{struggle to identify. Such limitations stem from the ambiguity of user queries and, more importantly, the complexity of comprehending large-scale and dynamic database values. Consequently, in real-world} scenarios where expert assistance is unavailable, existing methods suffer a severe performance collapse, with execution accuracy dropping by over 10\%. This underscores their lack of robustness due to the reliance on expert assistance.

To address this, we propose DIVER, a robust system that \textbf{\textit{automates}} evidence reasoning with dynamic interactive value linking. It leverages a \textit{compatible toolbox} containing diverse tools to probe the database. Then, restricted by a \textit{structured workspace (CoTF, Chain of Thoughts and Facts)}, it reflects based on probe results and selects a new tool for next round of probing. Through this automatically iterative process, DIVER identifies schema and value linking missed by existing methods. Based on these accurate linkings, DIVER is able to infer correct usage of SQL functions and formulas and generate high-quality evidence, achieving robust Text-to-SQL without expert assistance. Extensive experiments demonstrate that: 1) The DIVER system significantly enhances the robustness of various Text-to-SQL models, improving performance by up to \otheradd{10.82\%} in Execution Accuracy (EX) and \otheradd{16.09\%} in Valid Efficiency Score (VES). 2) Our dynamic interactive value linking \otheradd{significantly improves the robustness of existing systems and the accuracy of schema and value linking, especially when confronted with challenges posed by large-scale, dynamic database values.}
\end{abstract}




\received{July 2025}
\received[revised]{October 2025}
\received[accepted]{November 2025}

\maketitle

\section{Introduction}

Text-to-SQL aims to translate natural language questions (NLQ) into executable SQL queries on relational databases \cite{katsogiannis-meimarakisSurveyDeepLearning2023,qinSurveyTexttoSQLParsing2022,liuSurveyNL2SQLLarge2025}. It has always been regarded as an interface to databases, enabling non-technical users to query data using only natural language. As natural language serves as a core bridge in the current LLM-centric era, Text-to-SQL is playing an increasingly important role. More and more LLM-based agents are using Text-to-SQL interfaces to extract necessary information from databases. For instance, Business Intelligence (BI) and data analysis systems intensively leverage Text-to-SQL tools to query data and generate detailed analytical reports based on them . This places a higher demand on the full-process automation of the Text-to-SQL task \cite{xuDAgentRelationalDatabaseDriven2025, zhuChat2QueryZeroshotAutomatic2024, wuAutomatedDataVisualization2024, parikhAutoquerySimpleNatural2022}.

Real-world Text-to-SQL tasks often face challenges such as massive databases, dirty values, ambiguous user questions, and vague column names \cite{liCanLLMAlready2024, leiSpider20Evaluating2024, ganExploringUnderexploredLimitations2021, luomaSNAILSSchemaNaming2025}
. A new generation of Text-to-SQL benchmarks, represented by BIRD-bench~\cite{liCanLLMAlready2024}, is dedicated to constructing tasks closer to real-world scenarios, providing a solid data foundation for building practical natural language interfaces for databases. Numerous recent models have achieved remarkable progress on BIRD-bench. To improve Text-to-SQL accuracy, researchers have invested great amounts of effort to improve the underlying model architecture \cite{liCodeSBuildingOpensource2024, gaoPreviewXiYanSQLMultigenerator2025, pourrezaDINSQLDecomposedIncontext2023, gaoTexttoSQLEmpoweredLarge2023, qiRASATIntegratingRelational2022, wangRATSQLRelationAwareSchema2021, wangMACSQLMultiAgentCollaborative2024, pourrezaCHASESQLMultiPathReasoning2024, liDawnNaturalLanguage2024, maamariDeathSchemaLinking2024}.

However, we find that the impressive performance of existing models heavily relies on auxiliary of expert-written evidence. Specifically, different sizes of models all exhibit a significant performance \textbf{collapse} of over 10\% in Execution Accuracy (EX) when external expert evidence is unavailable. This reveals that existing Text-to-SQL systems face severe performance degradation in real-world scenarios, as non-technical users cannot provide expert-level insights, and LLM-based agents also cannot rely on human assistance with the demand for high automation. Unfortunately, this heavy reliance on expert-written evidence has been largely overlooked in recent research. According to the latest BIRD leaderboard, among the 52 methods, only 5 methods report performance without evidence in their papers or on the leaderboard. The state-of-the-art among these 5 methods ranks only 33rd on the leaderboard. Therefore, \textbf{how existing Text-to-SQL models maintain robustness in real-world scenarios without expert assistance} is a pressing issue to be addressed.

This paper aims to address the critical challenge of maintaining robust Text-to-SQL performance in real-world scenarios without relying on expert-written evidence.

\begin{figure}[tbp]
  \centering
  \includegraphics[width=\linewidth]{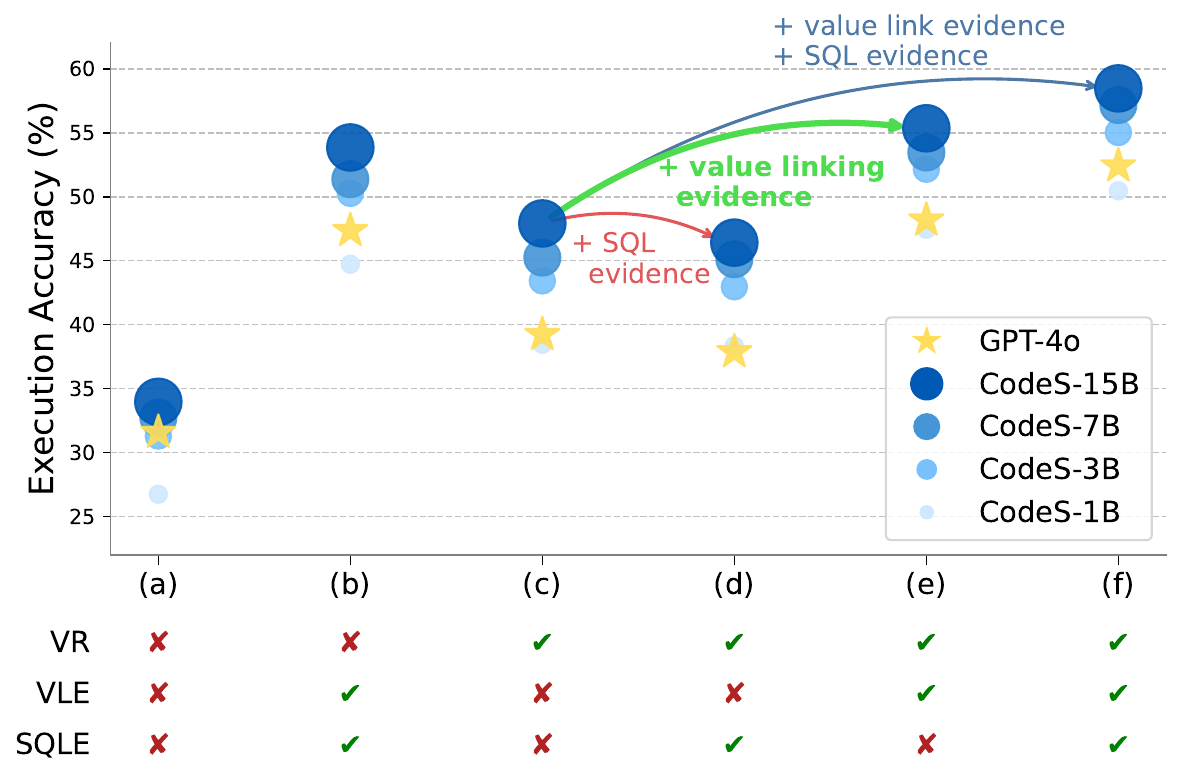}
  \caption{Fine-grained analysis of expert-written evidence reveals that effective \textbf{value linking} is a critical prerequisite for robust Text-to-SQL performance. "VR" indicates the Value Retrieval module of methods. "VLE" represents expert-written value linking and schema linking evidence, and "SQLE" represents expert-written SQL-related Evidence such as function usage.}
  \label{fig:intro-1}
\end{figure}

\textbf{The Key Bottleneck Behind the Performance Collapse.} To further explore the reason for such a collapse, we analyzed expert-written evidence on the BIRD-dev dataset, dividing it into VLE (value and schema linking evidence) and SQLE (SQL functions and formulas evidence). As shown in Figure~\ref{fig:intro-1}, a significant gap exists between current value retrieval methods and expert-written evidence (c $\rightarrow$ e). Besides, the impact of this gap is profound, as SQL-related evidence is beneficial only when the value evidence is correct (c $\rightarrow$ d decrease vs. c $\rightarrow$ f increase). \addrone{It is also important to note that this gap cannot be bridged by simply scaling up to larger LLMs such as GPT-4o. This is because the fundamental issue is not a deficiency in the model's reasoning capabilities, but rather a lack of access to complete, dynamic database values.} {Therefore, to achieve robust Text-to-SQL system without expert assistance, the key bottleneck lies in the incomplete value linking strategies.} {Current methods, relying on simple semantic similarity, often fail to handle ambiguous questions or non-semantic database content. For example, mapping “state special school” to its abbreviation “SSS” is difficult for existing methods.}

\begin{figure}[tbp]
  \centering
  \includegraphics[width=\linewidth]{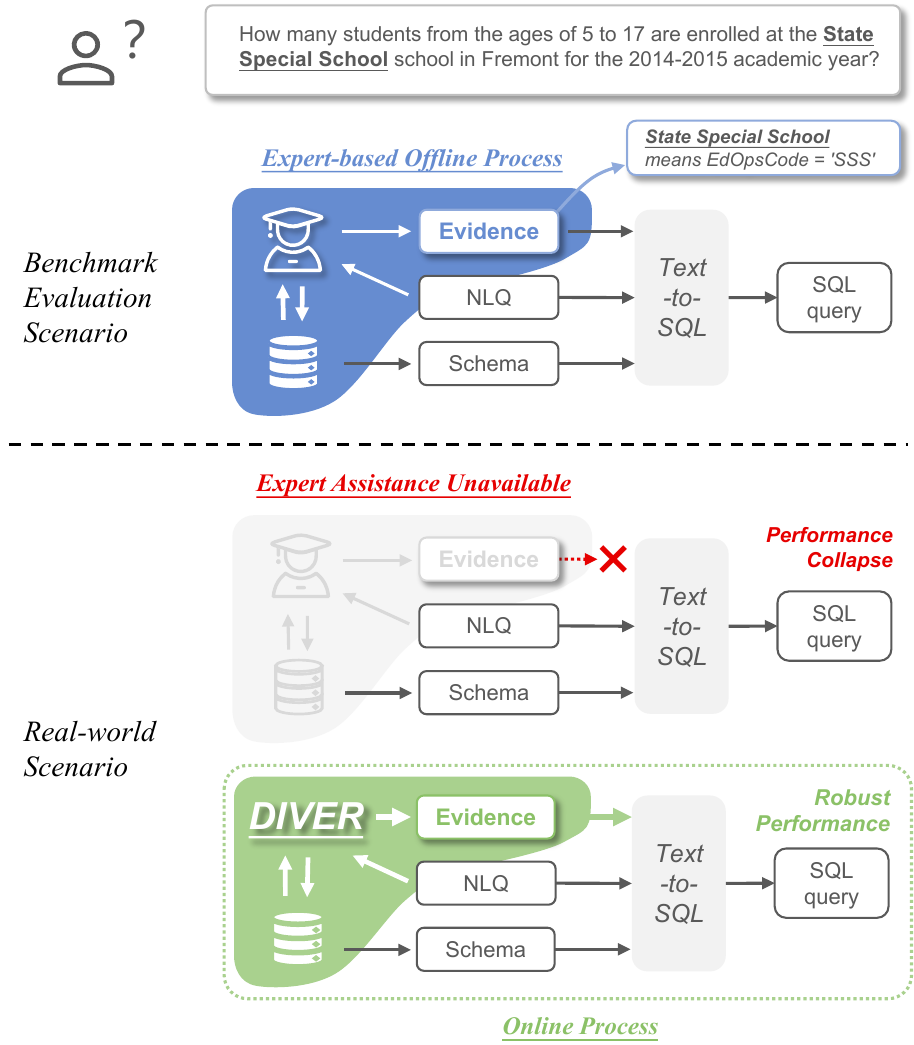}
  \caption{Existing Text-to-SQL methods utilize an offline expert-written evidence to aid the model, achieving high but unrealistic performance. They suffer a severe performance collapse in real-world scenario where expert is unavailable. DIVER system provides a training-free, online process that acts as an automated expert, generating evidence to achieve robust performance.}
  \label{fig:intro-case}
\end{figure}

To tackle these challenges, We propose DIVER, a novel, training-free system that automates high-quality evidence generation through \textbf{D}ynamic \textbf{I}nteractive \textbf{V}alue-linking and \textbf{E}vidence \textbf{R}easoning. DIVER mimics the iterative reasoning process of human experts using a multi-agent framework, comprising three key components: 1) the Break up Assistant decomposes complex NLQs into manageable sub-clauses for focused analysis. 2) The Look up Assistant conducts multi-turn interactive value linking and evidence reasoning with the database using a predefined toolbox in a structured Chain of Thoughts and Facts (CoTF) workspace. 3) The Evidence Assistant synthesizes structured information into natural language evidence tailored to different model preferences with style-aligned generation. This design enables DIVER to effectively handle ambiguous queries and large-scale values that are highly diverse and dynamic, thereby generating adaptive evidence that enhances the robustness of existing Text-to-SQL models. Through extensive experiments \otheradd{on 4 benchmarks and different scales of models}, we show that DIVER consistently boosts the accuracy of state-of-the-art models. On BIRD benchmark, DIVER improves up to \otheradd{10.82\%} in execution accuracy and \otheradd{16.09\%} in valid efficiency score.

Our contributions can be summarized as follows:

\begin{itemize}
  \item We identify and systematically analyze a critical but largely overlooked problem: the performance collapse of Text-to-SQL models in real-world scenarios due to their heavy reliance on expert-written evidence.
  \item We propose DIVER, a novel, training-free, and model-agnostic system that automates the generation of high-quality evidence by simulating an expert’s iterative reasoning and database exploration process.
  \item We introduce a core methodology of multi-turn interactive value linking, enabled by a structured Chain of Thoughts and Facts (CoTF) workspace and an extensible and compatible toolbox, which robustly grounds the reasoning of evidence in verified database facts.
  \item We demonstrate through comprehensive experiments that DIVER significantly enhances the robustness of various text-to-SQL models on 4 benchmarks.
\end{itemize}

\section{\addrone{Related Work}}

\paragraph{Text-to-SQL}
Text-to-SQL systems convert natural language queries (NLQs) to SQL for non-expert database access. Early approaches, such as RATSQL \cite{wangRATSQLRelationAwareSchema2021} and Global-GNN \cite{boginGlobalReasoningDatabase2019,boginRepresentingSchemaStructure2019}, used graph neural networks for schema-query modeling. Pre-trained models like T5 \cite{raffelExploringLimitsTransfer2023} advanced this via fine-tuning, as in RESDSQL \cite{liRESDSQLDecouplingSchema2023} and PICARD \cite{scholakPICARDParsingIncrementally2021}. Recent LLM-based methods, including DIN-SQL \cite{pourrezaDINSQLDecomposedIncontext2023}, DAIL-SQL \cite{gaoTexttoSQLEmpoweredLarge2023}, and SQL-specific LLMs \cite{liCodeSBuildingOpensource2024,liOmniSQLSynthesizingHighquality2025,gaoPreviewXiYanSQLMultigenerator2025}, leverage few-shot prompting and CoT reasoning. However, these models depend heavily on expert-written evidence, leading to significant accuracy drops (over 10\% in Execution Accuracy) without it.

\paragraph{Schema Linking and Value Linking}

Text-to-SQL converts natural language questions into executable SQL queries, a process that relies on schema linking (mapping question entities to database schema) and value linking (mapping them to actual database values). Early methods in the pre-trained language model (PLM) era mainly used explicit schema linking techniques such as relationally-aware embeddings \cite{wangRATSQLRelationAwareSchema2021}, while value linking received less attention \cite{guoComplexTexttoSQLCrossDomain2019, brunnerValueNetNaturalLanguagetoSQL2021}. With the rise of large language models (LLMs), the field shifted toward implicit linking, where models infer connections from rich schema information and retrieved data by semantic similarity matching \cite{sunSQLPaLMImprovedLarge2023, dongC3ZeroshotTexttoSQL2023}. However, they face challenges in retrieving non-semantic values such as abbreviations or numbers. More importantly, semantic similarity does not always reflect logical relevance. For instance, retrieving a single similar value might result in an incorrect WHERE column = 'value' clause, overlooking cases that require a LIKE 'value\%' pattern. These issues underscore the need for stronger value linking approaches that can handle the complexity and variability of dynamic, real-world databases.

\paragraph{Multi-agent Collaboration for Database}
Multi-agent frameworks are widely used in database tasks. Systems like DB-GPT \cite{boginGlobalReasoningDatabase2019}, DAgent \cite{xuDAgentRelationalDatabaseDriven2025}, and Chat2Query \cite{zhuChat2QueryZeroshotAutomatic2024} focus on Business Intelligence, treating Text-to-SQL as a black-box tool without enhancing its core process. Other methods target Text-to-SQL, using agents to decompose complex problems and boost accuracy \cite{wolfsonBreakItQuestion2020,pourrezaDTSSQLDecomposedTexttoSQL2024,liAlphaSQLZeroshotTexttoSQL2025,dengReFoRCETexttoSQLAgent2025,xieOpenSearchSQLEnhancingTexttoSQL2025,maPlugandplayNaturalLanguage2024,yangAutomatedValidatingFixing2025}. Yet, incomplete value linking leaves them prone to hallucinations with ambiguous data. Inspired by Embodied AI \cite{zhangBuildingCooperativeEmbodied2024,yaoReActSynergizingReasoning2023}, DIVER introduces a paradigm where agents interact with the environment to explore and validate, enhancing reasoning.

\section{Motivation}

\begin{figure}[tbp]
  \centering
  \includegraphics[width=\linewidth]{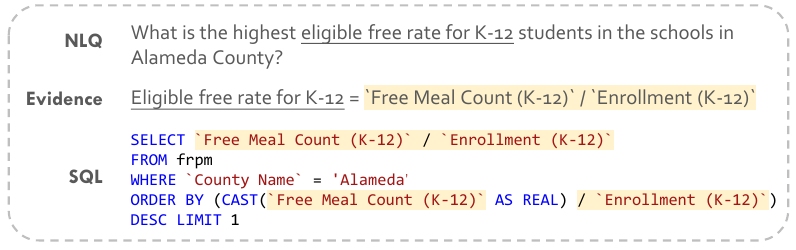}
  \caption{An Example of Evidence and Golden SQL Overlap.}
  \label{fig:overlap_example}
\end{figure}

\begin{figure}[tbp]
  \centering
  \includegraphics[width=\linewidth]{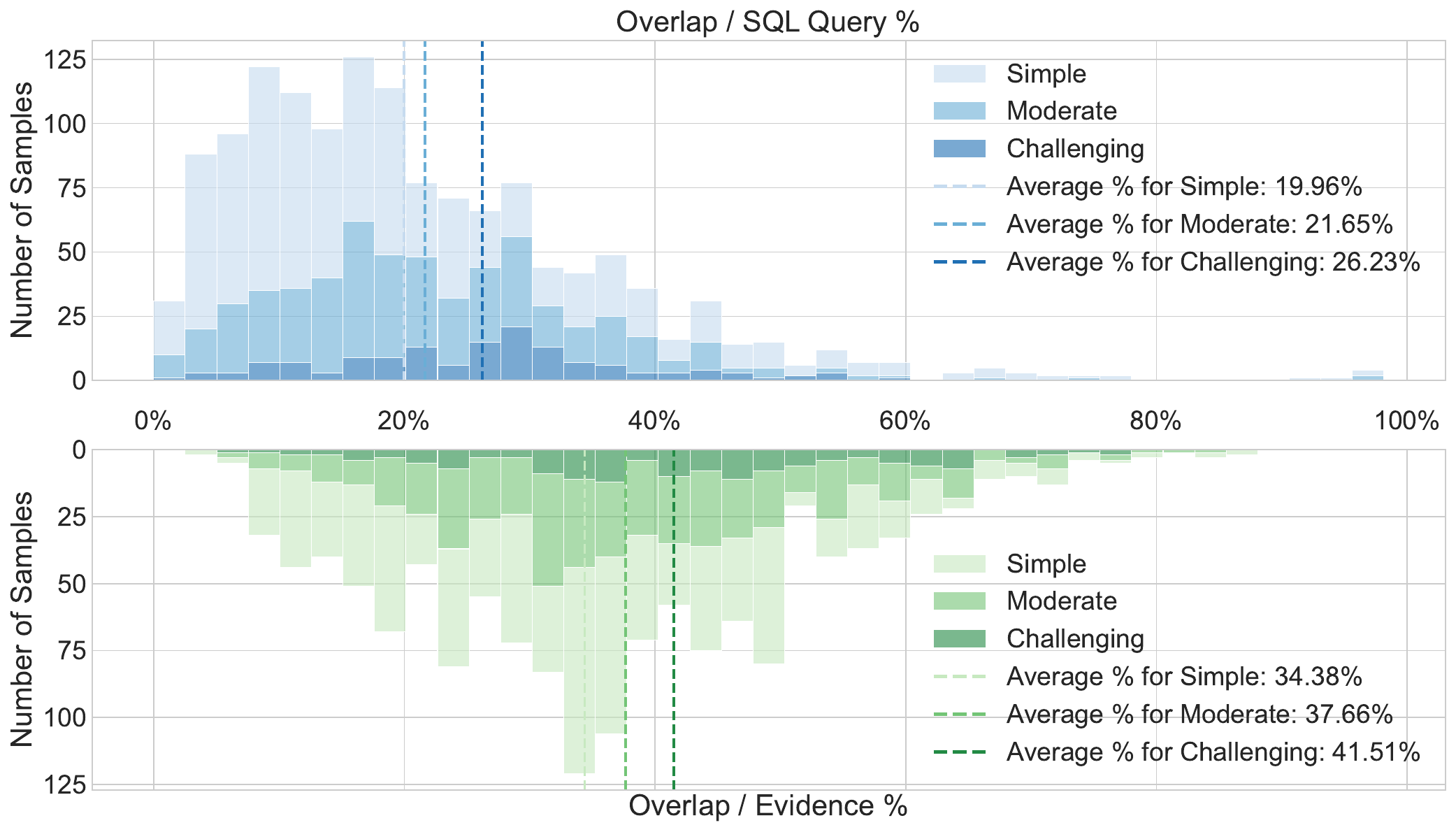}
  \caption{\addrthr{Distribution of token overlap between expert-written evidence and the golden SQL query on the BIRD-dev dataset. The green plot shows significant answer leakage within the expert-written evidence, while the blue plot shows the degree to which this leakage directly contributes to the final SQL query.}}
  \label{fig:dist_overlap}
\end{figure}

\begin{table}[tbp]
  \centering
  \caption{\addrthr{Performance on simulated real-world scenario that without answer leakage. Performance of models trained with evidence significantly drops, whereas models trained without evidence (Train w/o Evi.) exhibit greater robustness, which reveals unfair evaluation results with answer leakage. EX: Execution Accuracy; VES: Valid Efficiency Score.}}
  \scalebox{0.85}{\begin{tabular}{lccccccccc}
    \toprule
              & \multicolumn{4}{c}{Train w/ Evi.~$\rightarrow$~} & \multicolumn{4}{c}{Train w/o Evi.~$\rightarrow$~} \\
    \textbf{} & \multicolumn{2}{c}{{w/ Evi.}} & \multicolumn{2}{c}{{w/o Leak.}}  & \multicolumn{2}{c}{{w/o Evi.}} & \multicolumn{2}{c}{{w/o Leak.}}\\
    \cmidrule(lr){2-5}\cmidrule(lr){6-9}
    & EX\% & VES\% & EX\% & VES\% & EX\% & VES\% & EX\% & VES\%\\
    \midrule
    CodeS-1B & 50.46 & 51.07 & 34.81 & 40.67 & 38.46 & 41.77 & 37.61 & 46.23\\
    CodeS-3B & 55.02 & 56.54 & 40.35 & 44.46 & 43.42 & 44.55 & 42.57 & 46.34\\
    CodeS-7B & 57.17 & 58.80 & 41.72 & 45.85 & 45.24 & 48.13 & 44.98 & 51.71\\
    \bottomrule
  \end{tabular}}
  \label{tab:wo_overlap}
\end{table}

\begin{figure*}[tbp]
  \centering
  \includegraphics[width=0.93\textwidth]{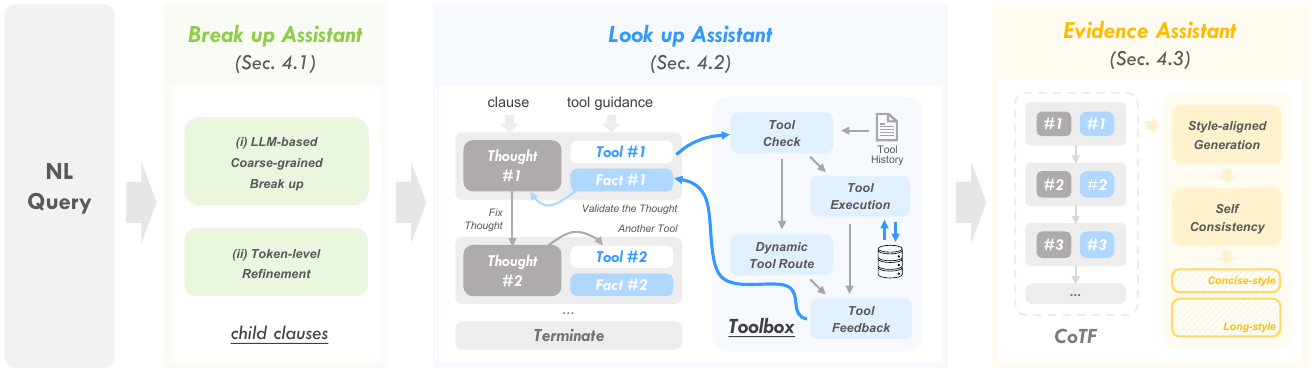}
  \caption{An Overview of DIVER Workflow.}
  \label{fig:pipeline}
\end{figure*}


\paragraph{The strength of this straightforward evidence.} 

As shown in Section 1, the expert-written evidence provided in the BIRD benchmark can significantly enhance model performance. This is because the evidence directly pinpoints complex schema and value links that are challenging for models to discover on their own. In the process of solving Text-to-SQL problems, experts engage in intensive reasoning and verification based on NLQ and database values, ultimately producing evidence that encapsulates critical and concise insights. It helps establish a \textbf{practical upper bound} and shows the level of performance achievable when a model is provided with all necessary information. With expert-written evidence, the task could be simplified to a more direct translation into SQL.

\paragraph{Limitations of Expert-written Evidence.}

However, for Text-to-SQL to be viable as a practical tool integrated into databases or employed by autonomous agents, relying on expert-written evidence raises questions about the generalizability and robustness. Moreover, the evidence is not conceptual guidance, but often has direct literally overlap with the golden SQL query, as shown in Figure. \ref{fig:overlap_example}. This is a form of \textbf{answer leakage} that should be noticed. More broadly, we have quantified the overlap on the BIRD-dev dataset as shwon in Figure \ref{fig:dist_overlap}, revealing the severe answer leakage between evidence and the golden SQL. Besides, as shown in Table \ref{tab:wo_overlap}, the answer leakage present in the evidence creates a significant gap between the training environment and real-world scenarios. Consequently, models exhibit a substantial performance drop when evaluated in a more realistic, leakage-free setting.

Except for the limitaion that expert-written evidence does not exist in real-world scenarios, there are still other limitations. For example, evidence that is set out for human understanding may not fit with the LLM's own way of thinking. Besides, expert-written evidence is static and cannot fits different models that need different kinds of guidance, which is demonstrated in our experiments. These limitations show the importance of an automated approach to generate high-quality evidence.

\section{System Overview}

To address the critical challenge of performance collapse in Text-to-SQL systems in real-world scenarios without expert assistance, we propose DIVER, a novel system for \textbf{D}ynamic \textbf{I}nteractive \textbf{V}alue-linking and \textbf{E}vidence \textbf{R}easoning. DIVER automates the generation of high-quality, model-adaptive evidence by mimicking the iterative and exploratory reasoning process of human experts. As shown in Figure~\ref{fig:pipeline}, DIVER adopts a multi-agent collaborative framework, consisting of the Break up Assistant, Look up Assistant, and Evidence Assistant.

The \textit{Break up Assistant} first segments the NLQ into a series of clauses that are semantically and lexically coherent\addrtwo{, each designed to contain only a single entity to be explored}. On one hand, this is crucial for complex NLQs with multiple intents, as it helps the Look up Assistant focus on \addrtwo{single entity} for deeper exploration. On the other hand, by decoupling the Break up Assistant with the Lookup Assistant, we address the risk of sub-problem "drifting" during subsequent multi-turn exploration.

The clauses segmented by the Break up Assistant are used to initialize the structured workspace of the Look up Assistant, which is structured along two dimensions: clauses and interaction turns. For each clause, the \textit{Look up Assistant} engages in multi-turn interactive value linking and evidence reasoning with the database using a predefined toolbox. The interaction process includes formulating thoughts and hypotheses (Thought), invoking appropriate tools, receiving feedback from the tools, validating or adjusting hypotheses based on the feedback, and invoking new tools to validate revised thoughts. Once the Look up Assistant uncovers solid factual evidence for each clause, the entire analysis process terminates, forming a structured, visual, and interpretable trace of reasoning and verification, which we refer to as the \textbf{Chain of Thoughts and Facts (CoTF)}.

Finally, the \textit{Evidence Assistant} synthesizes the structured information, which is grounded in solid evidence from the CoTF, into a final piece of natural language evidence. Through style alignment and self-consistency, it can generate evidence in different styles to align with the specific preferences of existing Text-to-SQL models. Existing models can directly capture the necessary information without additional alignment. For instance, SFT-based models that is constrained by input length may prefer shorter and concise evidence, while prompt-based methods may benefit from evidence with richer context for more detailed reasoning.

Another significant advantage of DIVER is that it is a \textit{training-free} system. It leverages the inherent capabilities of LLMs in reasoning and tool usage. This makes DIVER an out-of-the-box system that can be seamlessly integrated into various existing Text-to-SQL systems to enhance their robustness and practicality in real-world scenarios.

\section{System Design}

\begin{figure*}[tbp]
  \centering
  \includegraphics[width=\textwidth]{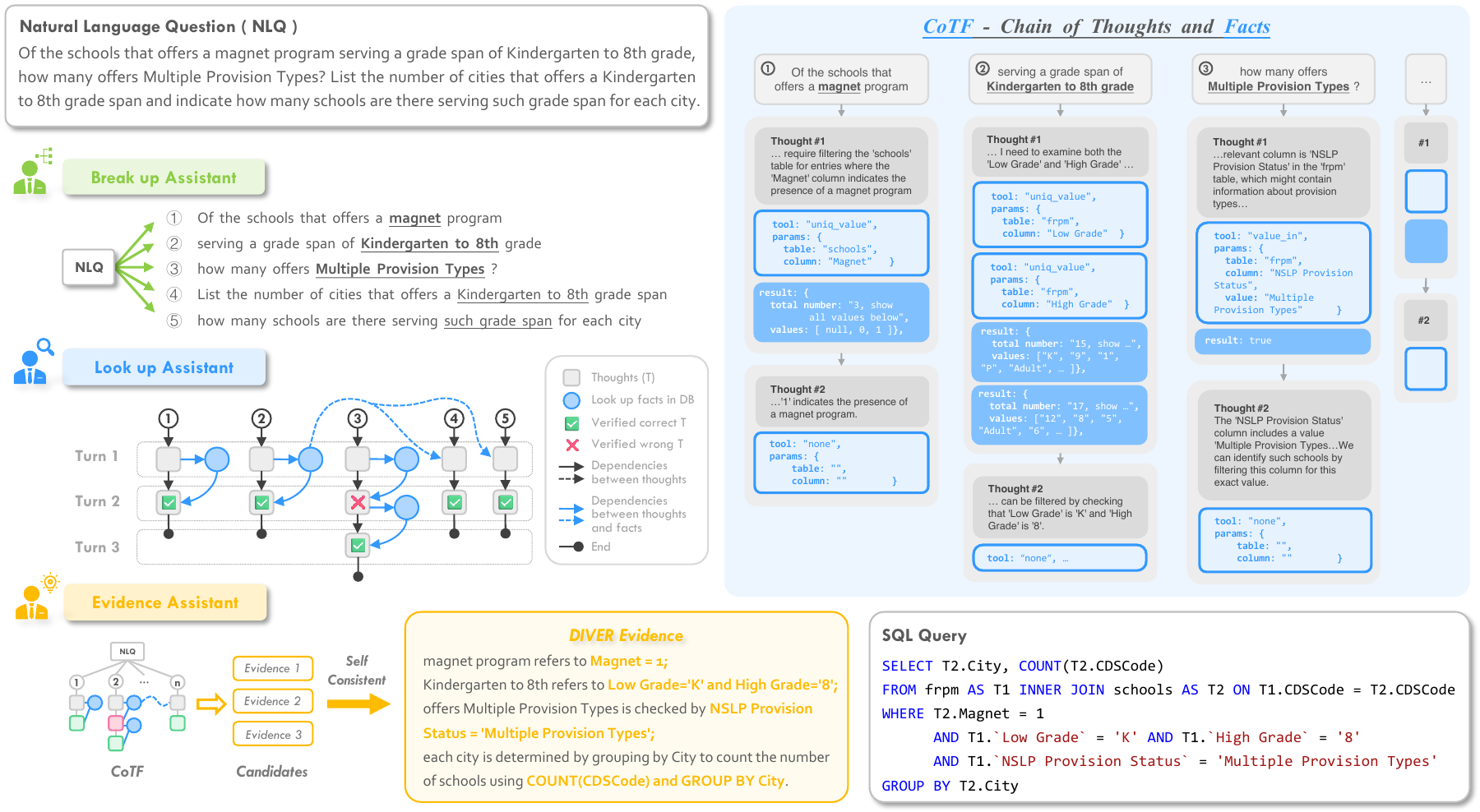}
  \caption{A Detailed Example of DIVER Workflow and CoTF.}
  \label{fig:example}
\end{figure*}


\subsection{Break up Assistant}
Real-world user queries often exhibit a high degree of complexity, which can manifest in diverse syntactic structures, multiple query intents, and semantic ambiguity. Directly addressing such complex queries cause models to lose focus and struggle to conduct in-depth exploration. Therefore, we employ a “divide-and-conquer” strategy: by decomposing the NLQ into a series of sub-queries, the Lookup Assistant is able to perform more targeted reasoning.

We design a \underline{\textbf{two-stage process}} to achieve this. First, an LLM performs a coarse-grained semantic segmentation, aiming for \textbf{each segment to correspond to a single SQL condition or entity}. Second, to ensure keeping the subtle semantics of NLQ, a token-level refinement module compares the segmentation result against the original NLQ word-by-word, \textbf{correcting any LLM-induced alterations or omissions}. The final segmentation is then used to initialize the Lookup Assistant’s structured workspace.

\subsection{Look up Assistant}
The Look up Assistant is the core component of the DIVER system, designed to mimic the \textbf{iterative and exploratory reasoning process} employed by human experts when facing complex problems. It dynamically discovers and validates value linking through multi-turn interactions with the database, ultimately generating a high-quality, fact-grounded reasoning chain for the subsequent Evidence Assistant.

\subsubsection{Chain of Thoughts and Facts: A Structured Workspace}
The Look up Assistant operates within a structured workspace we term the \textbf{Chain of Thoughts and Facts (CoTF)}. As illustrated in Figure~\ref{fig:example}, CoTF is a JSON object that organizes the reasoning process along two dimensions: the sub-clauses from the Break up Assistant and the multi-turn interactions for each clause. Each interaction step contains a thought, a tool called, and the tool feedback. In the \texttt{thought} section, the LLM can freely output with natural language. In the \texttt{tool\_called} section, the LLM is constrained with structured output to select a tool from our predefined toolbox.

Unlike the free-form and potentially ungrounded reasoning of a standard Chain of Thought (CoT), CoTF enforces a strict, verifiable loop:

\begin{enumerate}
  \item Grounded Reasoning: Every thought must be validated by a subsequent tool call that interacts with the database, ensuring the entire process is anchored in factual facts.
  \item Structured Actions: The tool-calling mechanism transforms abstract reasoning into a transparent and traceable sequence of actions, providing a solid evidence chain.
\end{enumerate}

The complete CoTF JSON object is passed back to the LLM after each turn. This provides the Look up Assistant with a global view, enabling information sharing between different sub-tasks. A Fact verified through a tool call in one clause can be directly utilized by others. For instance, in Figure~\ref{fig:example}, clauses $\textcircled{4}$ and $\textcircled{5}$ reference entities mentioned in preceding clauses. The Look up Assistant can directly leverage the verified information about the "magnet program" and "grade span" without re-invoking tools.

\subsubsection{Dynamic Interactive Value Linking: Thought-Verify-Refine}
Based on the CoTF structured workspace, we design a toolbox with various tools to allow the Look up Assistant to explore the database and performing \textbf{Dynamic Interactive Value Linking}. This process is a \textbf{iterative "dialogue"}. Specifically, for a semantic entity in the NLQ (e.g., "Kindergarten to 8th grade"), the LLM first forms an initial \textbf{thought or hypothesis}, such as, "This might correspond to values in the \texttt{Low Grade} and \texttt{High Grade} column." \addrthr{It then selects the most appropriate tool from the toolbox (e.g., \texttt{uniq\_value} or \texttt{sim\_value\_in}) to query the database and retrieve a \textbf{fact}, a raw value or schema item directly obtained from the database, without any correctness judgment about value linking and schema linking. The value linking or schema linking about this fact will be employed in the next thinking process: if the fact aligns with the user’s intent (e.g., finding the value `K' in the \texttt{Low Grade} column and the value '8' in the \texttt{High Grade} column), it is promoted to a \textbf{verified linking}, which is then summarized as part of the final evidence by evidence assistant. If the fact fails to support the hypothesis, the system uses the new information from this unsuccessful attempt to refine its thoughts, formulate a new hypothesis, and initiate the next round of interaction.}

\paragraph{\textbf{Exploring the Database via a Toolbox.}} Our tool design philosophy draws inspiration from the Embodied AI paradigm. In Embodied AI, an agent interacts with a physical environment through a series of atomic operations like "observe" and "grasp" to progressively understand the environment and complete tasks. This approach of decomposing complex tasks into atomic operations enhances system interpretability and extensibility. The Text-to-SQL task naturally possesses an "environment" to interact with (i.e., the database), and a universal interface (i.e., SQL). Unlike many open-domain QA tasks, the content retrieved from a database via SQL is \textbf{objective, truthful, and verifiable}, providing ideal conditions for constructing a reliable perception-action loop.

We designed a series of tools, each considered an "atomic operation" for interacting with the database, collectively defining the action space of the Look up Assistant. The benefits are twofold:
\begin{itemize}
    \item \textbf{Tool as a Guider:} Tools transform the ambiguous goal of "understanding the database", which is difficult to describe precisely with prompts, into a series of concrete, executable, and verifiable steps. By understanding and autonomously calling these tools, the LLM naturally accomplishes the exploration and comprehension of the database.
    \item \textbf{Tool as an Explainer:} The chain of tool calls clearly records the evolution of the model's thoughts, making the process of value linking and evidence generation transparent and interpretable. Compared to the highly unconstrained CoT, a chain of tool calls is also easier to analyze and intervene in.
\end{itemize}

\begin{figure}
  \centering
  \includegraphics[width=\columnwidth]{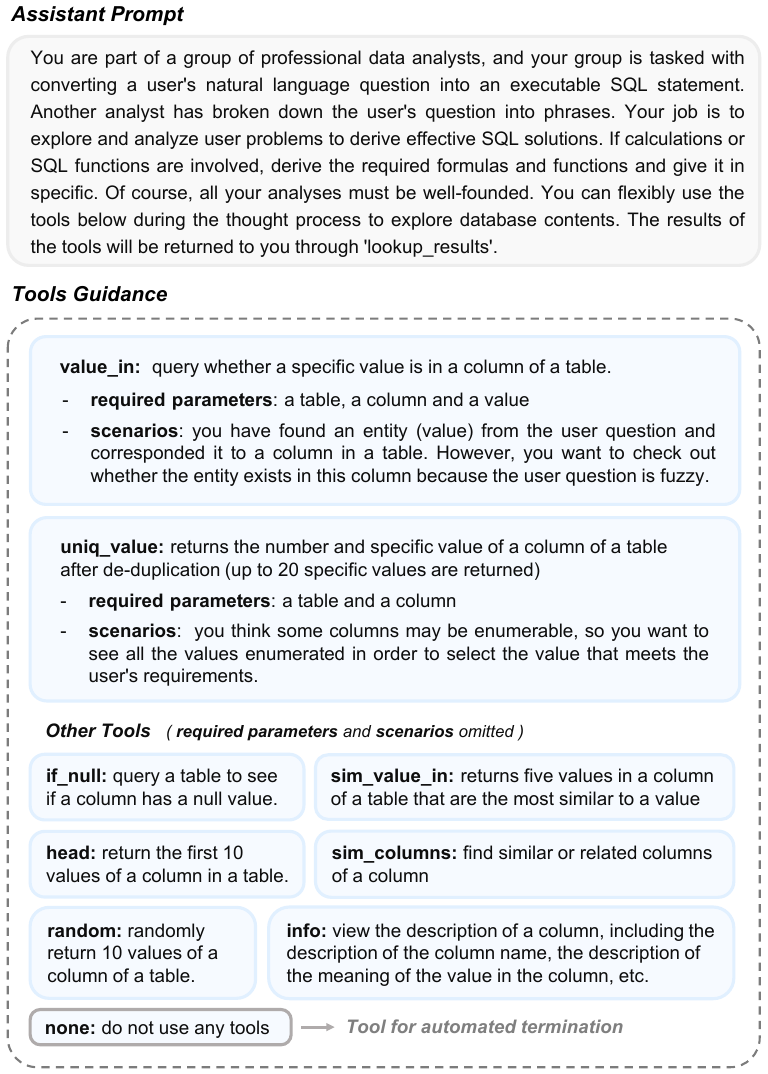}
  \caption{An Example of Lookup Assistant Prompt and Tool Guidance Prompt}
  \label{fig:tool_guidance}
\end{figure}

\addrtwo{Specifically, as shown in Figure~\ref{fig:tool_guidance}, we design eight tools covering three major functions: Value Probing (exact match, semantic similarity, and reverse linking from database values to NLQ), Data Inspection (sampling and null-value checks), and Schema Exploration (retrieving metadata and discovering related columns). For example, when a user asks about a “state special school”, a semantic similarity model may fail to match its abbreviation “SSS”. However, by inspecting the column content with \texttt{uniq\_value} and retrieving metadata by \texttt{info}, DIVER can discover “SSS” and infer their relevance.} As shown in Figure~\ref{fig:tool_guidance}, to enable the LLM to understand and correctly use these tools, we have meticulously designed three parts of guidance for each: a brief functional description, a list of required parameters, and a detailed description of applicable scenarios. In our ablation study, we validate the critical role of these three guidance components for system performance.

\paragraph{\textbf{Robust Self Correction via Tool Feedback}} Although the structured workspace and toolbox reduce the occurrence of LLM hallucinations, the model may still err when generating tool parameters (like table or column names) or get stuck in ineffective reasoning loops when facing complex problems. To address this, we designed an \textbf{in-loop feedback mechanism} that endows the system with powerful self-correction capabilities and robustness. This mechanism includes three types of feedback:
    \begin{itemize}
        \item \textbf{Standard Feedback:} When a tool is called for the first time and executes successfully, the system returns its execution result. In the next turn of thought, the LLM can use this new ``fact'' to confirm or refine its previous hypothesis.
        \item \textbf{Corrective Feedback:} When a tool call fails due to incorrect parameters (e.g., a non-existent table or column name), the system captures the exception and feeds the error message back to the LLM. This mechanism enables the system to automatically recover from common LLM hallucinations without complex hard-coded rules. According to our statistics on the BIRD-dev dataset, this mechanism successfully corrected 94 erroneous tool calls.
        \item \textbf{Guiding Feedback:} When the system detects that the model is repeatedly calling the same tool (with identical tool and parameters), it provides a ``Route Guide'' based on our predefined rules. For instance, for repeated calls to \texttt{value\_in}, the feedback might be, ``This tool has been called before. Consider using \texttt{sim\_value\_in} for a fuzzy search.'' Such repetitive behavior typically indicates that the model has entered an unproductive reasoning loop. Owing to the structured nature of CoTF, this state can be explicitly identified and and corresponding guidance can be provided. Since the suggestion is fed back as a contextual prompt, it preserves the assistant's decision-making freedom while effectively steering it out of the stagnated state.
    \end{itemize}


\subsection{Evidence Assistant}
After the Look up Assistant's process, the CoTF contains a wealth of accurate, verified, but highly structured information. However, existing Text-to-SQL models exhibit different preferences for the textual style of evidence due to variations in their model architectures and training details. For example, for a method like ChatGPT+CoT that relies on internal reasoning, overly concise evidence may degrade performance by mismatching the LLM's thought process, whereas evidence containing detailed analysis and verification might be more effective. Conversely, for models fine-tuned on the BIRD dataset (like \texttt{CodeS}), verbose evidence could lead to a performance drop due to distribution mismatch.

\begin{figure}
  \centering
  \includegraphics[width=\linewidth]{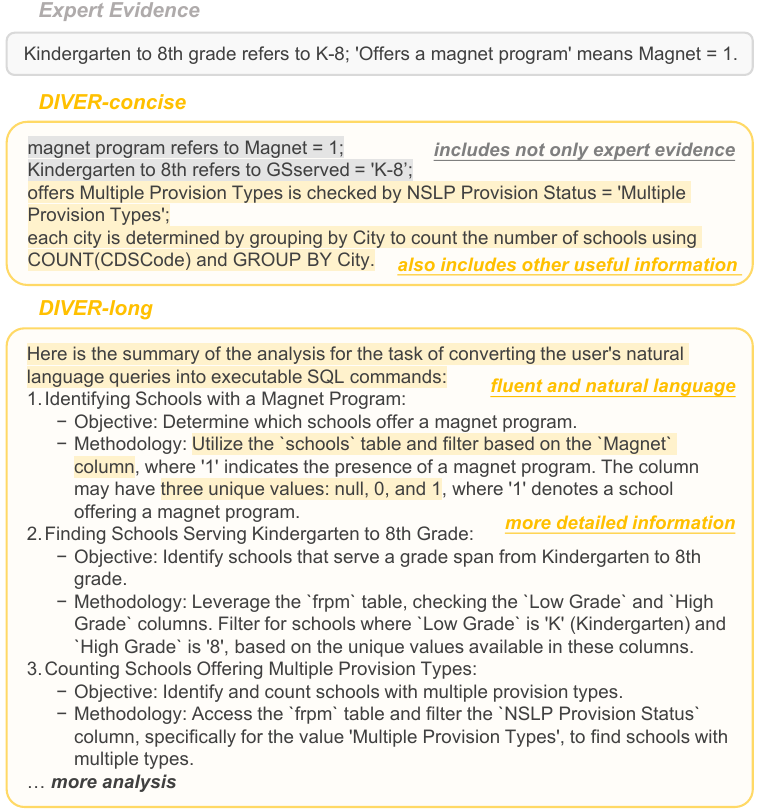}
  \caption{An Example of Different Evidence Styles of DIVER.}
  \label{fig:evidence_styles}
\end{figure}

Evidence Assistant is designed to \textbf{bridge the gap between the structured CoTF and the specific style of evidence required by the downstream Text-to-SQL model}. We designed two evidence generation styles, as shown in Figure~\ref{fig:evidence_styles}:
\begin{itemize}
    \item \textbf{DIVER-long:} A \textbf{long-form evidence} style that includes detailed analysis, thoughts, and verification processes. Its language is fluent and natural, making it more suitable for models that require rich context for reasoning.
    \item \textbf{DIVER-concise:} A \textbf{concise evidence} style that closely emulates the original expert-written evidence in BIRD, focusing on directly stating the core conclusions about value linking, schema linking, and function usage.
\end{itemize}

\paragraph{\textbf{Style-aligned Generation}}
To achieve style-aligned evidence generation, we employ a few-shot learning approach. In the Style-aligned Generation module, we provide the Evidence Assistant with five carefully constructed ``CoTF $\rightarrow$ concise evidence'' pairs as in-context examples. Since the LLM's task is primarily to learn and imitate the format and linguistic style of the concise evidence, rather than performing complex reasoning from scratch, a 5-shot context is sufficient for it to accurately generalize to new CoTF data.

\paragraph{\textbf{Self-Consistency}}
In the process of synthesizing the CoTF into natural language evidence, the LLM itself may produce hallucinations, such as omitting critical information from the CoTF or adding unsubstantiated details. Following common self-consistency strategies, we generate three different evidence. These three versions are then compared, and they supplement one another to produce a final, comprehensive version of the evidence.

\section{Experiments}

\newcolumntype{P}[1]{>{\centering\arraybackslash}p{#1}}
\begin{table*}[htbp]
    \centering
    \caption{
        Performance enhancement on SOTA Text-to-SQL models with DIVER system. ``Original'' denotes the baseline performance without BIRD evidence of each method. ``+ DIVER-long'' and ``+ DIVER-concise'' represent the results with two sytles of DIVER-generated evidence. \textcolor{blue}{Blue texts} indicate the specific improvement of DIVER over the baseline.
    }
    \label{tab:main_result}
    \scalebox{0.9}{\begin{tabular}{llP{1cm}P{1cm}P{2cm}P{2cm}P{2cm}P{2cm}}
      \toprule
      \multirow{2}{*}{\textbf{Method}} & \multirow{2}{*}{\textbf{Date}} & \multicolumn{2}{c}{\textbf{Original}} & \multicolumn{2}{c}{\textbf{+ DIVER-long}} & \multicolumn{2}{c}{\textbf{+ DIVER-concise}} \\
      \cmidrule(lr){3-4}
      \cmidrule(lr){5-6}
      \cmidrule(lr){7-8}
      & & \makecell[c]{EX\%} & \makecell[c]{VES\%} & \makecell[c]{EX\%} & \makecell[c]{VES\%} & \makecell[c]{EX\%} & \makecell[c]{VES\%} \\
      \midrule
      \multicolumn{8}{c}{\textit{with Close-source Foundation Models}}\\
      \midrule
      GPT-4o   & 2024-08 & 32.14 & 33.42 & 42.96 {\color{blue}\underline{(+10.82)}} & 49.51 {\color{blue}\underline{(+16.09)}} & 42.57 {\color{blue}\underline{(+10.43)}} & 44.58 {\color{blue}\underline{(+11.16)}} \\
      DIN-SQL + GPT-4o  & 2023-12 & 38.14 & 45.06 & 44.72 {\color{blue}(+6.58)}  & 50.11 {\color{blue}(+5.05)}  & 44.26 {\color{blue}(+6.12)}              & 51.24 {\color{blue}(+6.18)}  \\
      DAIL-SQL + GPT-4o & 2024-01 & 36.96 & 41.67 & 45.31 {\color{blue}(+8.35)}  & \textbf{\underline{54.30}} {\color{blue}(+12.63)} & \textbf{\underline{46.22}} {\color{blue}(+9.26)}              & 50.51 {\color{blue}(+8.84)}  \\
      \midrule
      \multicolumn{8}{c}{\textit{with Open-source Foundation Models}}\\
      \midrule
      SFT CodeS-1B & 2023-10 & 38.46 & 41.77 & 43.48 {\color{blue}(+5.02)} & 45.57 {\color{blue}(+3.80)} & 44.33 {\color{blue}\underline{(+5.87)}} & 48.76 {\color{blue}\underline{(+6.99)}} \\
      SFT CodeS-3B & 2023-10 & 43.42 & 44.55 & 46.81 {\color{blue}(+3.39)} & 48.11 {\color{blue}(+3.56)} & 48.63 {\color{blue}(+5.21)} & 49.93 {\color{blue}(+5.38)} \\
      SFT CodeS-7B & 2023-10 & 45.24 & 48.13 & 49.48 {\color{blue}(+4.24)} & 51.61 {\color{blue}(+3.48)} & 49.41 {\color{blue}(+4.17)} & 52.43 {\color{blue}(+4.3)} \\
      SFT CodeS-15B & 2023-10 & 47.91 & 49.60 & \underline{\textbf{50.91}} {\color{blue}(+3.00)} & \underline{\textbf{53.15}} {\color{blue}(+3.55)} & 49.74 {\color{blue}(+1.83)} & 52.96 {\color{blue}(+3.36)} \\
      XiYanSQL-QwenCoder-3B & 2025-04 & 35.27 & 38.78 & 43.02 {\color{blue}\underline{(+7.75)}} & 49.98 {\color{blue}\underline{(+11.2)}} & 39.11 {\color{blue}(+3.84)} & 44.18 {\color{blue}(+5.40)} \\
      XiYanSQL-QwenCoder-7B & 2025-04 & 39.63 & 40.32 & 43.35 {\color{blue}(+3.72)} & 48.90 {\color{blue}(+8.58)} & 41.33 {\color{blue}(+1.70)} & 41.44 {\color{blue}(+1.12)} \\
      XiYanSQL-QwenCoder-14B & 2025-04 & 41.33 & 41.95 & 46.02 {\color{blue}(+4.69)} & 49.78 {\color{blue}(+7.83)} & 44.26 {\color{blue}(+2.93)} & 47.66 {\color{blue}(+5.71)} \\
      \midrule
      \multicolumn{8}{c}{\textit{Code Unavailable Methods (just as a reference)}}\\
      \midrule
      ExSL + granite-20b-code & 2024-05 & 51.69 & 56.11 & - & - & - & - \\
      AskData + GPT-4o & 2025-03 & 65.91 & 63.34 & - & - & - & - \\
      \bottomrule
    \end{tabular}}
    \label{tab:main_results}
\end{table*}

\subsection{Experimental Setup}

\subsubsection{{Datasets.}} \otheradd{We conduct our main experiments on 4 text-to-SQL benchmarks, including two widely used datasets, BIRD \cite{liCanLLMAlready2024} and Spider \cite{yu-etal-2018-spider}, along with two more challenging datasets for robustness evaluation, DR.Spider \cite{chang2023drspider} and Spider-DK \cite{gan-etal-2021-spider-dk}. As DIVER is a training-free system, we directly evaluate it with the development sets of BIRD and Spider, containing 1,534 and 1,034 samples respectively. \textbf{\textit{Spider-dev}} comprises 166 databases spanning over 100 diverse domains, yet it is relatively basic and simple with small-scale database and clear user queries and database schemas. In contrast, \textbf{\textit{BIRD}} poses greater challenges, particularly in terms of database scale (each database contains on average around 549K rows, compared to Spider’s 2,000 rows), SQL function usage, and the presence of dirty values. \textbf{\textit{Spider-DK}} and \textbf{\textit{DR.Spider}} are derived from the Spider dataset by applying transformations and perturbations to the original samples, simulating real-world text-to-SQL scenarios in order to evaluate model's robustness. Spider-DK contains 535 transformed questions. DR.Spider is more challenging, featuring \textbf{17 carefully designed perturbation subsets} covering databases, user questions, and SQL queries, enabling a comprehensive assessment of model robustness across diverse conditions.}  

\subsubsection{Evaluation Mertics.} For all benchmarks, we apply \textbf{Execution accuracy (EX)} to evaluate whether the predicted and ground-truth SQL queries yield the same execution results on specific database instances. For the BIRD benchmark, \textbf{Valid Efficiency Score (VES)} is a special metric to evaluate SQL query performance by considering both accuracy and execution speed. Yet, preliminary experiments indicated that VES could be highly susceptible to fluctuations based on varying hardware, software, and system status. Hence, for BIRD, EX still serves as the stable and dependable metric.

\subsubsection{Models.} To evaluate the performance improvement brought by our DIVER system to existing Text-to-SQL methods, we selected the following SOTA (State-of-the-Art) models with various scales: the CodeS series (1B, 3B, 7B, 15B) \cite{liCodeSBuildingOpensource2024}, the XiYanSQL-QwenCoder series (3B, 7B, 14B) \cite{gaoPreviewXiYanSQLMultigenerator2025}, \otheradd{DIN-SQL \cite{pourrezaDINSQLDecomposedIncontext2023}, and DAIL-SQL \cite{dailsql}.}

\textbf{CodeS} and \textbf{XiYanSQL-QwenCoder} are both open-source models in the text-to-SQL domain, achieving SOTA among models of comparable sizes. \textbf{SFT-CodeS} is a version of the CodeS model that has been subsequently fine-tuned on the BIRD dataset or Spider dataset. We use the corresponding fine-tuned version directly to evaluate DIVER-generated evidence on BIRD-dev and Spider-dev. \otheradd{For Spider-DK and DR.Spider, we follow the experimental setup of the original paper \cite{liCodeSBuildingOpensource2024} and utilize the Spider-version SFT-CodeS for evaluation.} In contrast, XiYanSQL-QwenCoder is a in-context-learning-based model trained on mixed datasets, which could be utilized with prompt and few shots for all datasets. \otheradd{\textbf{DIN-SQL} and \textbf{DAIL-SQL} are widely compared methods built upon closed-source large-scale foundation models such as GPT-4. As GPT-4o now outperforms GPT-4 while being more cost-effective, we adopt GPT-4o as the foundation model for these two methods in our experiments.}

To utilize DIVER evidence in these models, we uniformly replaced the expert-written evidence from the BIRD dataset with the long-style or concise-style evidence generated by DIVER system. \otheradd{For Spider, Spider-DK, and DR.Spider, we directly concatenate the DIVER evidence to the user question. }For most models, their foundation models possess large context length, allowing them to directly process the long-style evidence. However, CodeS presents a special case. During its schema filtering, it utilizes the BERT model to retrieve relevant schemas with both NLQ and evidence. Since the length of long-style evidence exceeds BERT's context length limit, when evaluating long-style evidence on CodeS, we use only the NLQ for schema filtering and incorporated the DIVER-generated evidence during the SQL generation stage.

\definecolor{gain1}{HTML}{DEECF6} 
\definecolor{gain2}{HTML}{ADCCE9} 
\definecolor{gain3}{HTML}{73A9DB} 
\definecolor{gain4}{HTML}{2f75b5} 
\definecolor{badperf1}{HTML}{ffc9c9} 
\definecolor{badperf2}{HTML}{ff8f8f} 
\definecolor{badperf3}{HTML}{c00000} 
\definecolor{gold}{HTML}{FFD700}
\definecolor{silver}{HTML}{f56f28}
\definecolor{bronze}{HTML}{CD7F32}

\newcommand{\gain}[1]{\small\color{black}$\uparrow$ #1}
\newcommand{\loss}[1]{\small\color{gray}$\downarrow$ #1}
\newcommand{\eq}{\small\color{gray!80}$=$}

\newcommand{\gainone}[1]{\cellcolor{gain1}\gain{#1}}
\newcommand{\gaintwo}[1]{\cellcolor{gain2}\gain{#1}}
\newcommand{\gainthree}[1]{\cellcolor{gain3}\gain{#1}}
\newcommand{\gainfour}[1]{\cellcolor{gain4}\small\color{white}$\uparrow$ #1}
\newcommand{\badscore}[1]{\cellcolor{badperf1}#1}
\newcommand{\medscore}[1]{\cellcolor{badperf2}#1}
\newcommand{\vbadscore}[1]{\cellcolor{badperf3}\color{white}#1}

\newcommand{\goldmedal}{\color{gold}\faTrophy\color{black}}
\newcommand{\silvermedal}{\color{silver}\faMedal\color{black}}
\newcommand{\bronzemedal}{\color{bronze}\faAward\color{black}}
\newcommand{\first}[1]{#1\rlap{\raisebox{0.6ex}{\scriptsize\goldmedal\hspace{0.2em}}}}
\newcommand{\secnd}[1]{#1\rlap{\raisebox{0.6ex}{\scriptsize\silvermedal\hspace{0.2em}}}}
\newcommand{\thd}[1]{#1\rlap{\raisebox{0.6ex}{\scriptsize\bronzemedal\hspace{0.2em}}}}

\begin{table*}[!htbp]
\centering
\caption{Improvement of DIVER-concise on DR.Spider benchmark. For every perturbation (row), the top three performers are marked with \goldmedal, \silvermedal, and \bronzemedal \color{black} ~. For every method (column), red cells (\colorbox{badperf3}{\quad}, \colorbox{badperf2}{\quad}, \colorbox{badperf1}{\quad}) highlight three lowest scores.
The change after applying DIVER is shown with arrows ($\uparrow$,~$\downarrow$). 
Cell colors represent the magnitude of improvement: 
\colorbox{gain1}{\quad} for 0-2\%, 
\colorbox{gain2}{\quad} for 2-5\%, 
\colorbox{gain3}{\quad} for 5-10\%, and 
\colorbox{gain4}{\quad} for 10+\%. 
}
\label{tab:dr-spider}

\adjustbox{width=\textwidth, center}{%
\sisetup{detect-weight, mode=text, table-align-text-post=false}
\begin{tabular}{@{} l l S[table-format=4.0] S[table-format=2.1] S[table-format=2.1] S[table-format=2.1] S[table-format=2.1] c S[table-format=2.1] S[table-format=2.1] c S[table-format=2.1] S[table-format=2.1] c S[table-format=2.1] S[table-format=2.1] c @{}}
\toprule
\multirow{2}{*}{\textbf{Type}} & \multirow{2}{*}{\textbf{Perturbation}} & {\multirow{2}{*}{\textbf{\# Samples}}} & {\multirow{2}{*}{\small\begin{tabular}{@{}c@{}}\textbf{RESDSQL-3B}\\\textbf{+NatSQL}\end{tabular}}} & {\multirow{2}{*}{\small\begin{tabular}{@{}c@{}}\textbf{ChatGPT}\\\textbf{+ZeroNL2SQL}\end{tabular}}} & \multicolumn{3}{c}{\textbf{SFT-CodeS-1B}} & \multicolumn{3}{c}{\textbf{SFT-CodeS-3B}} & \multicolumn{3}{c}{\textbf{SFT-CodeS-7B}} & \multicolumn{3}{c}{\textbf{SFT-CodeS-15B}} \\
\cmidrule(lr){6-8} \cmidrule(lr){9-11} \cmidrule(lr){12-14} \cmidrule(lr){15-17}
& & {} & {} & {} & {EX\%} & \multicolumn{2}{c}{+ DIVER} & {EX\%} & \multicolumn{2}{c}{+ DIVER} & {EX\%} & \multicolumn{2}{c}{+ DIVER} & {EX\%} & \multicolumn{2}{c}{+ DIVER} \\
\midrule

\multirow{4}{*}{DB}
& schema-synonym        & 2619 & 68.3 & \thd{69.8} & 58.5 & \cellcolor{gain3}{65.7} & \gainthree{7.2} & \badscore{64.3} & \cellcolor{gain2}{69.2} & \gaintwo{4.9} & 67.2 & \cellcolor{gain3}{\secnd{73.9}} & \gainthree{6.7} & \badscore{66.9} & \cellcolor{gain3}{\first{74.0}} & \gainthree{7.1} \\
& schema-abbreviation   & 2853 & 70.0 & 74.8 & 68.6 & \cellcolor{gain2}{71.5} & \gaintwo{2.9}   & 75.0 & \cellcolor{gain2}{77.4} & \gaintwo{2.4} & 76.8 & \cellcolor{gain2}{\secnd{79.7}} & \gaintwo{2.9}   & \thd{78.7} & \cellcolor{gain1}{\first{80.2}} & \gainone{1.5}   \\
& Dbcontent-equivalence & 382  & \vbadscore{40.1} & \vbadscore{\thd{56.8}} & \medscore{53.9} & \cellcolor{gain2}{\first{58.4}} & \gaintwo{4.5}   & \vbadscore{47.9} & \cellcolor{gain3}{55.0} & \gainthree{7.1} & \vbadscore{46.9} & \cellcolor{gain4}\color{white}{\first{58.4}} & \gainfour{11.5} & \vbadscore{47.6} & \cellcolor{gain4}\color{white}{\secnd{57.6}} & \gainfour{10.0} \\
\cmidrule{2-17}
& Average               & {-}  & 59.4 & 67.1 & 60.3 & \cellcolor{gain2}{65.2} & \gaintwo{4.9}   & 62.4 & \cellcolor{gain2}{\thd{67.2}} & \gaintwo{4.8} & 63.6 & \cellcolor{gain3}{\first{70.7}} & \gainthree{7.1} & 64.4 & \cellcolor{gain3}{\secnd{70.6}} & \gainthree{6.2} \\
\midrule
\multirow{10}{*}{NLQ}
& keyword-synonym       & 953  & 72.4 & \first{74.0} & 60.9 & \cellcolor{gain2}{65.1} & \gaintwo{4.2}   & 70.9 & 70.4 & \loss{0.5} & \thd{73.0} & 72.6 & \loss{0.4} & \secnd{73.5} & 72.2 & \loss{1.3} \\
& keyword-carrier       & 399  & 83.5 & 88.2 & 86.5 & \cellcolor{gain2}{88.5} & \gaintwo{2.0}   & \thd{91.2} & 88.5 & \loss{2.7} & \secnd{91.5} & 90.2 & \loss{1.3} & \first{91.7} & 89.2 & \loss{2.5} \\
& column-synonym        & 563  & 63.1 & \medscore{62.7} & \badscore{56.0} & \cellcolor{gain2}{58.1} & \gaintwo{2.1}   & \medscore{60.0} & \cellcolor{gain2}{\secnd{63.8}} & \gaintwo{3.8} & \badscore{63.2} & \cellcolor{gain1}{\secnd{63.8}} & \gainone{0.6} & \medscore{\first{64.7}} & \thd{63.6} & \loss{1.1} \\
& column-carrier        & 579  & 63.9 & 71.7 & 67.4 & \cellcolor{gain1}{68.4} & \gainone{1.0}   & 74.4 & 73.8 & \loss{0.7} & \first{80.7} & \secnd{80.0} & \loss{0.7} & 79.1 & \cellcolor{gain1}{\thd{79.3}} & \gainone{0.2} \\
& column-attribute      & 119  & 71.4 & 70.6 & \vbadscore{47.9} & \cellcolor{gain4}\color{white}{64.7} & \gainfour{16.8} & 67.2 & \cellcolor{gain3}{\thd{73.1}} & \gainthree{5.9} & \medscore{63.0} & \cellcolor{gain4}\color{white}{\secnd{77.3}} & \gainfour{14.3} & 68.9 & \cellcolor{gain4}\color{white}{\first{79.0}} & \gainfour{10.1} \\
& column-value          & 304  & \first{76.6} & \thd{76.0} & 72.4 & 71.1 & \loss{1.3}   & 75.0 & 71.4 & \loss{3.6} & 73.7 & 72.0 & \loss{1.7} & \secnd{76.3} & 75.3 & \loss{1.0} \\
& value-synonym         & 506  & \medscore{53.2} & 70.6 & 59.7 & \cellcolor{gain3}{66.2} & \gainthree{6.5}   & 67.0 & 66.8 & \loss{0.2} & \first{72.7} & 71.2 & \loss{1.6} & \thd{71.9} & \cellcolor{gain1}{\secnd{72.5}} & \gainone{0.6} \\
& multitype             & 1351 & \badscore{60.7} & \badscore{66.4} & 57.5 & \cellcolor{gain2}{61.2} & \gaintwo{3.7}   & 66.5 & 65.8 & \loss{0.7} & \first{69.5} & \thd{68.7} & \loss{0.8} & \secnd{69.4} & 67.8 & \loss{1.6} \\
& others                & 2819 & 79.0 & 79.4 & 74.9 & 74.7 & \loss{0.2}   & 78.5 & 78.0 & \loss{0.5} & \first{81.5} & \thd{80.1} & \loss{1.4} & \secnd{81.2} & 80.0 & \loss{1.2} \\
\cmidrule{2-17}
& Average               & {-}  & 69.3 & 73.2 & 64.8 & \cellcolor{gain2}{68.7} & \gaintwo{3.9}   & 72.3 & \cellcolor{gain1}{72.4} & \gainone{0.1} & 74.3 & \cellcolor{gain1}{\thd{75.1}} & \gainone{0.8} & \secnd{75.2} & \cellcolor{gain1}{\first{75.4}} & \gainone{0.2} \\
\midrule
\multirow{6}{*}{SQL}
& comparison            & 178  & \first{82.0} & 73.6 & 60.7 & \cellcolor{gain3}{70.2} & \gainthree{9.5} & 69.7 & \cellcolor{gain2}{72.5} & \gaintwo{2.8} & \secnd{77.5} & \thd{74.7} & \loss{2.8} & 71.9 & 71.9 & \eq \\
& sort-order            & 192  & \first{85.4} & 80.2 & 69.8 & \cellcolor{gain2}{72.9} & \gaintwo{3.1}   & 79.2 & \cellcolor{gain2}{81.3} & \gaintwo{2.1} & 81.8 & \cellcolor{gain2}{\thd{83.9}} & \gaintwo{2.1} & \secnd{84.9} & \thd{83.9} & \loss{1.0} \\
& nonDB-number          & 131  & 85.5 & \first{92.4} & 84.7 & 84.0 & \loss{0.7}   & 87.8 & 84.7 & \loss{3.1} & \secnd{90.1} & \thd{88.6} & \loss{1.6} & 84.0 & \cellcolor{gain2}{87.8} & \gaintwo{3.8} \\
& DB-text               & 911  & 74.3 & \first{80.7} & 67.1 & \cellcolor{gain3}{74.1} & \gainthree{7.0}   & 77.2 & 76.6 & \loss{0.6} & \secnd{80.5} & \thd{80.4} & \loss{0.2} & \first{80.7} & 79.9 & \loss{0.8} \\
& DB-number             & 410  & \first{88.8} & \thd{86.1} & 80.5 & 76.6 & \loss{3.9}   & 85.1 & \cellcolor{gain1}{\thd{86.1}} & \gainone{1.0} & 84.9 & \cellcolor{gain1}{\secnd{86.8}} & \gainone{1.9} & 85.9 & 85.6 & \loss{0.3} \\
\cmidrule{2-17}
& Average               & {-}  & \first{83.2} & 82.6 & 72.6 & \cellcolor{gain2}{75.6} & \gaintwo{3.0}   & 79.8 & \cellcolor{gain1}{80.2} & \gainone{0.4} & \secnd{83.0} & \thd{82.9} & \loss{0.1} & 81.5 & \cellcolor{gain1}{81.8} & \gainone{0.3} \\
\midrule
\textbf{All} & \textbf{Global average} & {-} & 71.7 & 74.9 & 66.3 & \cellcolor{gain2}{70.1} & \gaintwo{3.8}   & 72.8 & \cellcolor{gain1}{73.8} & \gainone{1.0} & 75.0 & \cellcolor{gain1}{\first{76.6}} & \gainone{1.6} & \thd{75.1} & \cellcolor{gain1}{\secnd{76.5}} & \gainone{1.4} \\
\bottomrule
\end{tabular}
}
\end{table*}

\newcolumntype{G}{>{\columncolor{gray!20}}S[table-format=2.1]}
\renewcommand{\gain}[1]{\small\color{blue}$\uparrow$ #1}

\begin{table*}[htbp]
\centering
\caption{Performance comparison on Spider-dev and Spider-DK with DIVER-long and DIVER-concise.}
\label{tab:spider_dev_dk}
\scalebox{0.9}{\begin{tabular}{l G r l r l G r l r l}
\toprule
\multirow{2}{*}{\textbf{Method}} & \multicolumn{5}{c}{\textbf{Spider-dev}} & \multicolumn{5}{c}{\textbf{Spider-DK}} \\
\cmidrule(lr){2-6} \cmidrule(lr){7-11}
& {Original} & \multicolumn{2}{c}{+ DIVER-long} & \multicolumn{2}{c}{+ DIVER-concise} & {Original} & \multicolumn{2}{c}{+ DIVER-long} & \multicolumn{2}{c}{+ DIVER-concise} \\
\midrule

DIN-SQL + GPT-4o & 82.9 & 79.4      & \loss{3.5} & 82.6 & \loss{0.3} & 55.9 & 69.3 & \gain{13.4} & 66.0 & \gain{10.1} \\
DAIL-SQL + GPT-4o & 83.3 & 80.8     & \loss{2.5} & 82.9 & \loss{0.4} & 69.7 & 71.6 & \gain{1.9}  & 72.8 & \gain{3.1}\\
\midrule
XiYanSQL-QwenCoder-3B & 82.8 & 79.4 & \loss{3.4} & 80.5 & \loss{2.3} & 72.3 & 74.2 & \gain{1.9}  & 73.8 & \gain{1.5} \\
SFT CodeS-1B & 77.9 & 79.0          & \gain{1.1} & 78.9 & \gain{1.0} & 64.7 & 69.0 & \gain{4.3}  & 68.8 & \gain{4.1} \\
SFT CodeS-3B & 83.4 & 83.4          & \eq        & 83.4 & \eq        & 71.8 & 76.1 & \gain{4.3}  & 73.6 & \gain{1.8} \\
SFT CodeS-7B & 85.4 & 84.6          & \loss{0.8} & 84.9 & \loss{o.5} & 72.0 & 76.6 & \gain{4.6}  & 75.0 & \gain{3.0} \\
SFT CodeS-15B & 84.9 & 83.8         & \loss{1.1} & 85.0 & \gain{0.1} & 70.7 & 74.8 & \gain{4.1}  & 74.4 & \gain{3.7} \\
\bottomrule
\end{tabular}}
\end{table*}

\subsubsection{Implementation Details.} For all 4 benchmarks, SQLite serves as the database engine for hosting and management. However, we design and implement DIVER system in a compatible manner. It can be adapted to different SQL engines with rewriting less than 100 lines of code.

The specific version of the model we have chosen for DIVER is \texttt{gpt-4o-2024-08-06}. For the \texttt{sim\_value\_in} and \texttt{sim\_columns} tools, we employ the BERT model to assess semantic similarity. For the Breakup, Lookup, and Evidence Assistants, we set the temperature to 0.2, 0.7, and 0.7, respectively. During the dynamic interactive value linking process, we set the maximum number of turns to 5, at which point the exploration is forcibly terminated to prevent infinite loops. For the Evidence Assistant, we use 5 shots for the style alignment module and 3 candidates for self-consistency.

All experiments are employed on a server equipped with 1 NVIDIA A100 (80G) GPU, 1 NVIDIA A800 (80G) GPU, 40 Intel Xeon Silver 4210R CPUs, and the Ubuntu 22.04.1 operating system.

\subsection{Performance Evaluation}

\subsubsection{Accuracy Improvement on BIRD}


In this section, we conducted experiments on the BIRD-dev and report two metrics: Execution Accuracy (EX) and Valid Execution Score (VES). For each baseline model, we report its original performance without evidence assistance and the enhanced performance after applying two variants of DIVER-generated evidence, \textbf{DIVER-long} and \textbf{DIVER-concise}. The results are summarized in Table~\ref{tab:main_results}. DIVER provides significant improvements across all models. Key observations include:

\begin{enumerate}
    \item \addrone{\textbf{Consistent Gains Across Scales:} DIVER-long and DIVER-concise deliver consistent performance gains for models ranging from small-scale (e.g., CodeS-1B) to large-scale closed-source SOTA models (e.g., GPT-4o),  with a maximum improvement of 10.82\%.}

    \item \textbf{Style-specific Strengths:} DIVER is capable of automatically generating different forms of evidence tailored to different models. Experimental results show that DIVER-long achieves superior performance for prompt-based models as well as larger-scale models (e.g., CodeS-15B), as these models can leverage the richer information contained in long-style evidence. In contrast, DIVER-concise is more suitable for smaller-scale models (e.g., CodeS-1B), which typically exhibit lower inherent generalization ability and higher sensitivity to format alignment.
\end{enumerate}

\subsubsection{Robustness Improvement}

\addrone{In this section, we evaluate the DIVER's improvement on robustness on three datasets: Spider-dev, Spider-DK, and DR.Spider. Table \ref{tab:spider_dev_dk} and Table \ref{tab:dr-spider} show the results. In summary, while DIVER is less impactful for over-simplified datasets like Spider-dev, it brings significant robustness improvements under complex perturbations in Spider-DK and DR.Spider, with marked benefits for both large and small models.

Due to the Spider-dev's simplicity such as small databases, simple schemas, and direct value linking, most baseline models already achieve high accuracy on it (e.g., SFT-CodeS-15B: 84.9\%), leaving little room for improvement. Therefore, introducing additional DIVER evidence to Spider-dev occasionally causes slight performance drops, suggesting that it may become unnecessary or even distracting to apply DIVER to overly simple environments.

In contrast, baseline models face severe accuracy drops on Spider-DK and DR.Spider due to more complex variations and perturbations. For example, DB perturbations such as renaming a column or changing field values poses challenge for existing value-linking strategies. As shown in Table \ref{tab:dr-spider}, SFT-CodeS-15B only achieves 47.6\% of EX under DBcontent-equivalence perturbation of DR.Spider, revealing that current text-to-SQL systems struggle with dynamic database in real-world settings.

However, DIVER yields large improvements on difficult perturbations. Across models, DIVER sets new state-of-the-art scores for multiple perturbations, especially in DB-level perturbation, confirming the effectiveness of the dynamic interactive value linking in handling dynamic databases. Furthermore, for small-scale models like SFT-CodeS-1B that inherently lack robustness and generalization, DIVER strongly improves its performance of global average from only 66.3\% to 70.1\% (Table \ref{tab:dr-spider}), narrowing the gap with much larger 3B models.

While minor drops occur in a few perturbations where baselines are already strong, DIVER consistently improves the average performance, demonstrating its own robustness that it provides helpful information without introducing excessive noise.}

\subsubsection{Dynamic Interactive Value Linking vs. Existing Value Linking}

To conduct a detailed comparison between the dynamic interactive value linking in the DIVER system and existing methods, we designed a rigorous evaluation focusing on linking accuracy. For each sample, we extract schema items (tables and columns) and values from both the ground-truth SQL and the predicted SQL, forming a golden set of entities and a predicted set of entities. To ensure the accuracy of this extraction, we utilize sqlparse library to parse SQL queries. We then compute the F1-score for both schema linking and value linking as follows: \textbf{Precision} is the fraction of correctly predicted entities out of all predicted entities ($\frac{|\text{Golden Set} \cap \text{Predicted Set}|}{|\text{Predicted Set}|}$), measuring the accuracy of the predictions. \textbf{Recall} is the fraction of correctly predicted entities out of all entities in the golden set ($\frac{|\text{Golden Set} \cap \text{Predicted Set}|}{|\text{Golden Set}|}$), measuring the completeness of the predictions. The \textbf{F1-score} is the harmonic mean of these two metrics ($2 \times \frac{\text{Precision} \times \text{Recall}}{\text{Precision} + \text{Recall}}$), providing a single, balanced measure of performance, which is reported finally.

We compare DIVER against two representative schema linking and value linking techniques:

\begin{itemize}
  \item CodeS - Schema Filter and Value Retriever: Uses a classifier for schema pruning and a "coarse-to-fine" BM25+LCS search for values.
  \item XiYanSQL - M-Schema: Employs LLM-based retrieval for keywords and column selection via few-shot pruning.
  \item \addrthr{CodeS + Unique Values: an additional baseline to further investigate the role of value-related evidence. Our abaltion study shows that uniq\_value tool plays an important role, therefore, in this setup, we augment the CodeS retriever with up to 20 unique values for each retrieved column, following DIVER’s uniq\_value setup.}
\end{itemize}

\begin{table}[t]
\centering
\caption{Comparison of F1-Scores for schema and value linking across different methods and difficulty levels. The best results are \underline{\textbf{emphasized}}.}
\label{tab:linking_f1_performance}
    \scalebox{0.8}{\begin{tabular}{l c c c c c c c c}
    \toprule
    \multirow{2}{*}{\textbf{Method}} & \multicolumn{2}{c}{\textbf{Simple}} & \multicolumn{2}{c}{\textbf{Moderate}} & \multicolumn{2}{c}{\textbf{Challenging}} & \multicolumn{2}{c}{\textbf{Overall}} \\
    \cmidrule(lr){2-3} \cmidrule(lr){4-5} \cmidrule(lr){6-7} \cmidrule(lr){8-9}
    & \textbf{\footnotesize Schema} & \textbf{\footnotesize Value} & \textbf{\footnotesize Schema} & \textbf{\footnotesize Value} & \textbf{\footnotesize Schema} & \textbf{\footnotesize Value} & \textbf{\footnotesize Schema} & \textbf{\footnotesize Value} \\
    \midrule
    DIVER       & \underline{\textbf{88.06}} & \underline{\textbf{84.29}} & \underline{\textbf{84.89}} & 72.13 & \underline{\textbf{83.22}} & \underline{\textbf{72.87}} & \underline{\textbf{86.64}} & \underline{\textbf{79.53}} \\
    CodeS    & 87.23 & 84.03 & 83.66 & \underline{\textbf{73.21}} & 80.26 & 67.58 & 85.49 & 79.21 \\
    \quad + uniq                          & 78.88 & 82.65 & 77.63 & 72.54 & 75.56 & 66.58 & 78.19 & 78.07 \\
    XiYanSQL & 80.74 & 75.42 & 75.14 & 62.13 & 72.69 & 64.53 & 78.28 & 70.37 \\

    \bottomrule
    \end{tabular}}
\end{table}

\begin{figure}
  \centering
  \includegraphics[width=\columnwidth]{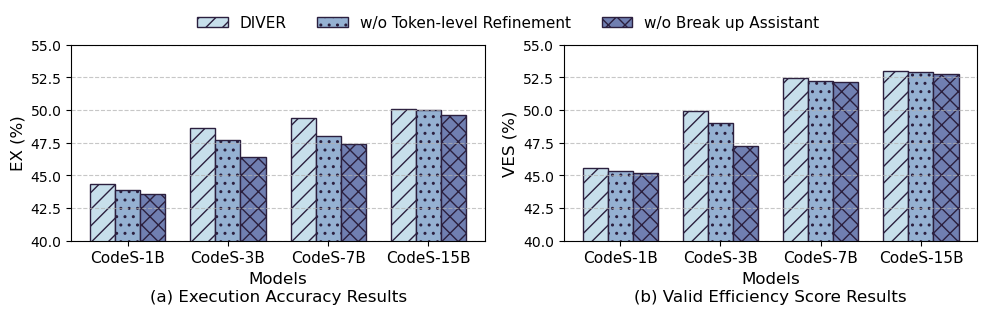}
  \caption{Abaltion Results of Break up Assistant. DIVER-concise is utilized in this experiment.}
  \label{fig:ablation_breakup}
\end{figure}

For DIVER and CodeS, we utilize the SFT-CodeS-3B as the base Text-to-SQL model. For XiYanSQL, we utilize the XiYanSQL-QwenCoder-3B. The experiment is employ on the BIRD-dev dataset. All SQL queries were generated in an expert-free setting, without access to expert-written evidence.

As shown in \ref{tab:linking_f1_performance}, the experimental results fully demonstrate the effectiveness of dynamic interactive value linking. The results show that DIVER’s interactive value linking surpasses existing methods in both schema and value linking accuracy in most cases, with only a slight decline in value linking on “moderate” questions.

Notably, DIVER achieves a particularly significant advantage on "challenging" questions. This is because challenging problems in BIRD often feature ambiguous user queries alongside highly complex data structures, where values and schema are difficult to comprehend even with description. These makes it difficulty for methods that solely rely on semantic similarity. In contrast, DIVER’s interactive approach allows it to explore the database dynamically and deeply understand both the ambiguous user intent and the complex data formats.

\addrthr{As for CodeS + Unique Values, the augmented unique values not only failed to close the gap with DIVER but also degraded CodeS’s performance, especially on schema linking. This stems from two factors: (1) the retrieval module of CodeS cannot exploit unique values to filter ambiguous or semantically weak column names due to context length constraints, and (2) large amounts of irrelevant unique values, especially from non-categorical columns (e.g., IDs), introduce noise that impact both linking and SQL generation. This underscores DIVER’s core advantage: rather than injecting more and more values, it distills them into concise, noise-free evidence that directly benefits SQL generation.}

\subsection{Ablation Study}

To analyze the contribution of each key component within \textbf{DIVER} system, we conduct a thorough ablation study. To ensure experimental efficiency while maintaining result reliability, experiments in ablation study were performed on a smaller, representative subset of the BIRD-dev, which we refer to as \texttt{small-dev}. This subset was constructed via stratified sampling 10\% data from the full BIRD-dev set, preserving the original data distribution with respect to question difficulty and different database. Besides, we employ SFT-CodeS-3B and DIVER-concise for the ablation experiments, unless stated otherwise. We organize the ablation study of the DIVER system around its three main components: the Breakup Assistant, the Lookup Assistant, and the Evidence Assistant.

\subsubsection{Ablation of Breakup Assistant}

Here, we quantify the impact of two key components: the \textbf{Breakup Assistant} as a whole, and its internal \textbf{Token-level Refinement} module. We compare the performance of the complete DIVER system against two ablated versions: (1) \textbf{w/o Break up Assistant}, where we bypass the decomposition process and use the original, complete NLQ in all subsequent steps. This means the Lookup Assistant now process the entire NLQ, which is more complex. So we set the maximum number of turns to 10 to ensure the process is not terminated before all necessary information can be explored. For the Evidence Assistant, we constructs 5 style-alignment shots based on the complete NLQ to ensure the style-alignment performance; (2) \textbf{w/o Token-level Refinement}, where we use the sub-clauses generated by the LLM without postprocessing. This means the sub-clauses may contain errors such as omissions or unintended alterations of the original NLQ. The experiments are conducted on the \texttt{SFT-CodeS} model family, from 1B to 15B, to assess the impact across different model scales.

Figure~\ref{fig:ablation_breakup} presents the results, both ablated versions lead to a noticeable degradation in performance across all model sizes. The removal of the Break up Assistant causes the most significant performance drop, \addrtwo{reaching up to over 2\%, which represents almost half of the total gains brought by full DIVER}. This underscores the critical importance of the NLQ decomposition stage. The Token-level Refinement module also proves to be a valuable contributor. Without it, LLM hallucinations can cause the generated sub-clauses to omit or alter words from the original user question. By applying token-level refinement, the hallucination issue is effectively solved, resulting in a corresponding improvement in performance.

\subsubsection{Ablation of Lookup Assistant}

The Lookup Assistant is central to DIVER's ability to ground its reasoning in the database's actual content. In this section, we conduct two sets of ablation studies to analyze its key components, including tools, route guidance, tool prompt, and CoTF.

\noindent\underline{\textbf{Ablation of Toolbox.}} Toolbox is the core component that enables Lookup Assistant to interact with the database. It includes 8 tools with prompts for description, parameters, and usage scenarios, plus a dynamic routing mechanism to avoid unproductive reasoning loops. Ablation study analyzes these components’ impacts. The \texttt{none} tool, which serves as a stop signal for the assistant to terminate exploration, is not considered as an exploration tool thus excluded from this ablation study.

We examine tool calling counts across experiments to detect substitutions between tools. By linking changes in usage and performance, we assess each tool’s significance and irreplaceability.



The results are presented in Table~\ref{tab:tool_ablation}. {\textit{Ablation of single tool}} shows the impact of removing each tool individually, including removing its prompt and removing it from callable tools. We observe that removing any single tool leads to a degradation in both EX and VES scores, demonstrating the effectiveness of the Lookup Assistant. This suggests the difficulty of understanding the database with single value retrieval strategy. Moreover, the varying frequencies of tool calling highlight the autonomous decision-making capability of the Lookup Assistant. Compared to static combinations of tools for information retrieval, dynamic and self-directed tool calling provides more precise information.

The most critical tools appear to be \textbf{\texttt{sim\_columns}} and \textbf{\texttt{uniq\_value}}. Removing \texttt{uniq\_value} results in the largest performance drop in {EX} (from 48.70 to 44.81), underscoring the importance of reverse value linking. Through \texttt{uniq\_value}, the assistant gains insight into the values contained within a column, enabling it to map these values back to entities mentioned in the natural language query (NLQ). This is particularly important when the user intent is ambiguous or database values lack sufficient semantic. Besides, we observe a notable increase in the calling of tools such as \texttt{value\_in}, \texttt{head}, \texttt{random}, and \texttt{if\_null}, indicating that the assistant attempts to compensate by gathering more information through alternative tools, which also leads to an increase in interaction turns (from 2.38 to 2.67). Nevertheless, both EX and VES scores still drop significantly, further emphasizing the irreplaceable role of \texttt{uniq\_value}.

Similarly, removing \texttt{sim\_columns} leads to the same degradation in EX score, demonstrating its central role in the DIVER system. As discussed in the Method section, when the assistant lacks a clear plan for the next step, it often falls into inefficient reasoning loops—such as repeatedly querying the same column even when no useful information is returned. In addition to using the tool router to guide reasoning, \texttt{sim\_columns} serves as a guidance tool itself. Based on the idea of "tool as a guider", its presence reminds the assistant to switch columns for querying. When \texttt{sim\_columns} is removed, the frequencies of other tools do not significantly increase, possibly because the assistant becomes trapped in a reasoning loop, and chooses to terminate the process after receiving repeated calling feedback about redundant tool use.

Other tools also reveal valuable insights. For example, ablating \texttt{value\_in} has limited impact on EX and VES, as tools like \texttt{sim\_value\_in} can almost entirely substitute it in value retrieval. Nevertheless, we retain \texttt{value\_in} in the DIVER system because it is more efficient and lightweight compared to \texttt{sim\_value\_in}. Although \texttt{sim\_value\_in} has its limitations, it remains indispensable as a fuzzy search tool that supports broader value retrieval, especially when column values are not enumerable (e.g., usernames or geographic names). The removal of \texttt{if\_null} also results in a noticeable performance drop. We observe an increase in \texttt{value\_in} calling (+11), likely as a partial compensation. However, this number still falls short compared to the original 36 calling of \texttt{if\_null}, suggesting that the replacement is inadequate. This again illustrates the importance of tool itself that serve as guides in the reasoning process.

Overall, our results confirm the effectiveness of leveraging tools as both guiders and explainers. Furthermore, they show that each tool is crucial for complete understanding of database values, requiring multiple retrieval strategies to work together in a complementary fashion.

\begin{table}[tbp]
  \centering
  \caption{Ablation study of the CoTF (Chain of Thoughts and Facts). The bar chart shows the distribution of counts of calling each tool. The tools, from left to right, are: \ding{192}~\texttt{value\_in}, \ding{193}~\texttt{sim\_value\_in}, \ding{194}~\texttt{uniq\_value}, \ding{195}~\texttt{head}, \ding{196}~\texttt{random}, \ding{197}~\texttt{if\_null}, \ding{198}~\texttt{info}, and \ding{199}~\texttt{sim\_columns}.}
  \label{tab:cotf_ablation}
  \begin{tabular}{m{0.4cm}<{\centering}m{0.4cm}<{\centering}m{3cm}<{\centering}m{1cm}<{\centering}m{0.5cm}<{\centering}m{0.5cm}<{\centering}}
      \toprule
      \multicolumn{2}{c}{\textbf{Components}} & \multirow{2}{*}{\textbf{\#Tool Calling}} & \multirow{2}{*}{\textbf{\#Turns}} & \multirow{2}{*}{\textbf{EX\%}} & \multirow{2}{*}{\textbf{VES\%}} \\
      \cmidrule{1-2}
      \textbf{CoT} & \textbf{CoF} \\
      \midrule
      \ding{55} & \ding{55} & - & - & 30.52 & 31.17 \\
      \ding{51} & \ding{55} & - & - & 32.47 & 33.12 \\
      \ding{55} & \ding{51} & \includegraphics[width=\imagewidth]{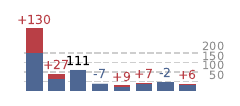} & 3.31 & 47.40 & 48.37 \\
      \ding{51} & \ding{51} & \includegraphics[width=\imagewidth]{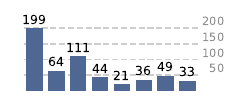}     & 2.38 & 48.70 & 50.49 \\
      \bottomrule
  \end{tabular}
\end{table}

\begin{table}[tbp]
  \centering
  \caption{Ablation study of the toolbox. See Table\ref{tab:cotf_ablation} for the description of the bar chart.}
  \label{tab:tool_ablation}
  \begin{tabular}{m{1.5cm}m{3cm}<{\centering}m{1cm}<{\centering}m{0.6cm}<{\centering}m{0.6cm}<{\centering}}
      \toprule
        & \textbf{\#Tool Calling} & \textbf{\#Turns} & \textbf{EX\%} & \textbf{VES\%} \\
      \midrule
      DIVER & \includegraphics[width=\imagewidth]{fig/bar_plot/plot_our.pdf}                      & 2.38 & 48.70 & 50.49 \\
      \midrule
      \multicolumn{5}{c}{\textit{Ablation of single tool}}\\
      \midrule
      w/o value\_in & \includegraphics[width=\imagewidth]{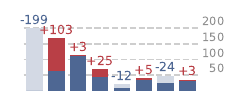}          & 2.57 & 48.05 & 50.15 \\
      w/o sim\_value\_in & \includegraphics[width=\imagewidth]{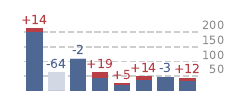} & 2.81 & 47.40 & 48.11 \\
      w/o uniq\_value & \includegraphics[width=\imagewidth]{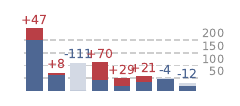}      & 2.67 & 44.81 & 46.73 \\
      w/o head & \includegraphics[width=\imagewidth]{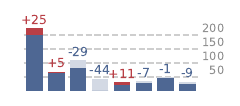}                   & 2.66 & 45.45 & 46.91 \\
      \makecell[l]{w/o\\random} & \includegraphics[width=\imagewidth]{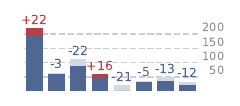}               & 2.70 & 48.05 & 48.45 \\
      \makecell[l]{w/o\\if\_null} & \includegraphics[width=\imagewidth]{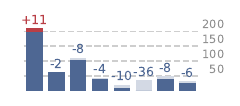}            & 2.62 & 46.75 & 48.03 \\
      w/o info & \includegraphics[width=\imagewidth]{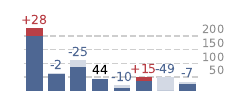}                   & 2.70 & 46.10 & 46.36 \\
      w/o sim\_columns & \includegraphics[width=\imagewidth]{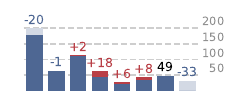}    & 2.68 & 44.81 & 46.32 \\
      \midrule
      \multicolumn{5}{c}{\textit{Ablation of tool route and prompt}}\\
      \midrule
      \makecell[l]{w/o\\tool route} & \includegraphics[width=\imagewidth]{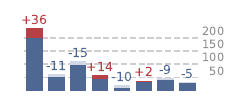}       & 2.68 & 48.70 & 51.95 \\
      \makecell[l]{w/o\\description} & \includegraphics[width=\imagewidth]{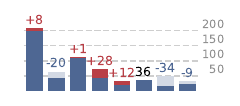}       & 2.76 & 48.05 & 48.98 \\
      \makecell[l]{w/o\\parameter} & \includegraphics[width=\imagewidth]{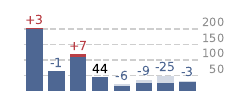}            & 2.50 & 45.45 & 46.74\\
      \makecell[l]{w/o\\scenario} & \includegraphics[width=\imagewidth]{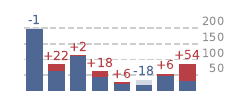}        & 2.21 & 44.81 & 47.98\\
      \bottomrule
  \end{tabular}
\end{table}

{\textit{Ablation of tool route and prompt}} in Table~\ref{tab:tool_ablation} shows the impact of route mechanism and each aspect of tool prompt. Disabling \texttt{tool route} does not decrease the EX score and even increase the {VES} to 51.95. However, it results in an increase in calling \texttt{value\_in} and decrease other calling of tools, showing a more concentrated distribution of tool-calling. The most severe degradation comes from removing the \texttt{scenario} examples, underscoring the importance of it for guiding the complex tool-use behavior. We can also observe that, the calling of tools becomes more more uniform, indicating the assistant’s insufficient understanding of the tools.

\noindent\underline{\textbf{Ablation of the CoTF.}} {Chain of Thought and Fact (CoTF)} integrates LLM reasoning (\textit{Thought}) with grounded evidence from tool use (\textit{Fact}). To validate the effectiveness of CoTF, we compare four distinct configurations as shown in Table~\ref{tab:cotf_ablation}.

\begin{itemize}[topsep=0pt]
    \item \textbf{Full DIVER (\texttt{CoT} $\checkmark$, \texttt{CoF} $\checkmark$):} Our proposed system, which combines reasoning and tool-based fact-checking. It achieves the highest performance (48.70\% \texttt{EX}, 50.49\% \texttt{VES}) with the greatest efficiency (2.38 average turns).

    \item \textbf{w/o CoT (\texttt{CoT} $\boldsymbol{\times}$, \texttt{CoF} $\checkmark$):} In this setting, we disable the model's ability to output its reasoning chain ("thought") and force it to directly select a tool. The performance drops significantly. Crucially, the model resorts to a less efficient, brute-force exploration strategy, evidenced by the sharp increase in both total tool calls (+130) and average turns (3.31). This demonstrates that the reasoning step is vital for formulating an efficient and effective exploration plan.

    \item \textbf{w/o CoF (\texttt{CoT} $\checkmark$, \texttt{CoF} $\boldsymbol{\times}$):} This configuration represents a traditional Chain-of-Thought approach where the model can reason but cannot use tools to verify facts or explore the database. While it can generate a plausible-sounding plan, it cannot ground its reasoning in the actual database content, often leading to unexecutable SQL due to factual inaccuracies (e.g., hallucinating column or value names).

    \item \textbf{Baseline (\texttt{CoT} $\boldsymbol{\times}$, \texttt{CoF} $\boldsymbol{\times}$):} This is a standard text-to-SQL baseline where the model generates SQL directly from the NLQ and schema, guided by a few-shot prompt to ensure output format compliance. This approach lacks any interactive exploration of facts and deep thinking, achieving the lowest performance.
\end{itemize}

\begin{figure*}
  \centering
  \includegraphics[width=\linewidth]{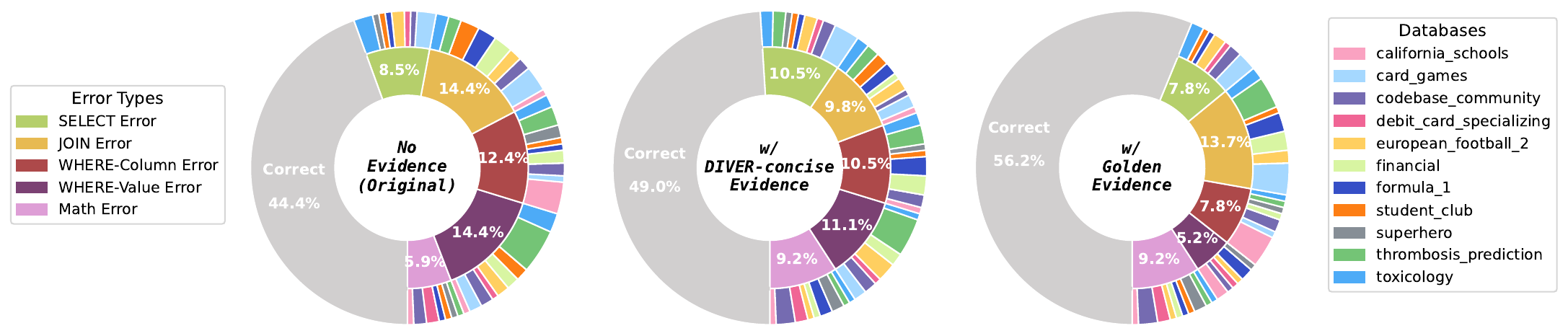}
  \caption{Distribution of error types for baseline model, DIVER-concise, and upper bound.}
  \label{fig:error_ana}
\end{figure*}

\begin{table}[t]
  \centering
  \caption{Ablation study of components in Evidence Assistant (EA). ''SA'' denotes style-alignment module.}
  \label{tab:ablation_evidence}
    \scalebox{0.97}{\begin{tabular}{llcccccc}
    \toprule
    \textbf{} & \textbf{Format} &  \textbf{\#Avg Length} & \textbf{EX\%} & \textbf{VES\%} \\
    \midrule
    \textbf{DIVER-long}      & NL          &  373 & \textbf{49.35}  & \textbf{52.02}  \\
    w/o EA                   & json        &  763 & 43.51  & 45.09  \\
    \midrule
    \textbf{{DIVER-concise}} & NL          &  71  & \textbf{48.70} & \textbf{50.49} \\
    w/o SA                   & NL          &  124 & 45.45 & 46.59 \\
    \multirow{2}{*}{w/o EA}  & NL template &  322 & 41.56 & 43.40 \\
                             & json        &  174 & 42.21 & 44.27 \\
    \bottomrule
    \end{tabular}}
\end{table}

In summary, the results powerfully demonstrate the synergistic relationship between reasoning and fact-checking. Removing either component cripples the system: without thought (\texttt{CoT}), exploration becomes inefficient and reactive; without facts (\texttt{CoF}), reasoning becomes ungrounded and prone to hallucination. Our CoTF architecture successfully leverages both to achieve superior performance and efficiency.

\subsubsection{Ablation of Evidence Assistant}



Evidence Assistant aims to synthesize structured CoTF into different styles of summary, providing high-quality, contextualized evidence for the final Text-to-SQL model. To validate the effectiveness of it, we conducted a comprehensive ablation study, with results presented in Table~\ref{tab:ablation_evidence}.

We conducted ablation studies for two target evidence styles: long and concise. For the long-style evidence, we design a setting without the Evidence Assistant (DIVER-long w/o EA), where the JSON-formatted CoTF is passed directly to the Text-to-SQL model.

For the concise-style evidence, the DIVER system includes a style-alignment module (SA) designed to ensure that the generated evidence aligns with the expert-written evidence. To ablate the SA module, we prompt the Evidence Assistant to generate evidence as concise as possible (DIVER-concise w/o SA). Besides, we also ablate the Evidence Assistant for concise-style evidence. We explored two alternative approaches. First, leveraging the structured nature of CoTF, we design concise natural language (NL) templates for each tool. For example, the template for the \texttt{uniq\_value} tool is \texttt("Table ‘{table}’ column ‘{column}’ has {total} unique values, samples: {', '.join(values)}."), where table and column are parameters for calling the tool, and total and values are the results returned by the tool. Second, we directly extracted the "facts" section from CoTF, which consists of the JSON for all tool calls and their return results. It is important to note that this setup differs from the w/o CoT setting in the ablation study of CoTF. In this setting, the Lookup Assistant is allowed to output the CoT part, but we only use the CoF to generate SQL, whereas w/o CoT does not allow the Lookup Assistant to output the CoT part at all.

The experimental results demonstrate the effectiveness of the Evidence Assistant. For the long style evidence, summarizing the CoTF into natural language yields significantly more compact evidence, reducing the length by half compared to the raw JSON CoTF. Furthermore, the raw JSON CoTF is not limited to necessary information. It also including noise from trial-and-error steps, presenting a greater challenge for the Text-to-SQL model, leading to a sharp decline in performance.

For the concise style evidence, the style-alignment module achieves the most streamlined evidence (average length of only 71) while maintaining high quality. In contrast, prompting the model to be concise is unstable, resulting in a longer average length (124) and a slight drop in both EX and VES scores. As for the two w/o EA settings, they suffer from a similar problem as the DIVER-long w/o EA setting: neither the NL template nor the raw fact JSON can filter out irrelevant content from the CoTF. Even with a carefully designed concise NL template, the average length still reached 322, as it included substantial noise that negatively impacted performance.

\section{\addrtwo{Error Analysis}}

\addrtwo{To further understand the strengths and limitations of DIVER, we conducted a detailed error analysis on BIRD-small-dev set used in ablation study, with 154 samples in total. As illustrated in Figure \ref{fig:error_ana}, the analysis compares the error distribution of the baseline model, the model augmented with DIVER-concise evidence, and an upper-bound performance with manually crafted golden evidence.

The primary observation is that DIVER significantly reduces the most critical and frequent errors. The baseline model's failures are concentrated in JOIN errors (14.4\%) and WHERE-Value errors (14.4\%). With DIVER's assistance, JOIN errors are substantially cut down to 9.8\%, and WHERE-Value errors are reduced to 11.1\%. This demonstrates DIVER's effectiveness in providing crucial evidence for establishing complex table relationships and performing accurate value linking, leading to an overall accuracy boost from 44.4\% to 49.0\%.

However, a performance gap still exists between DIVER and the Golden Evidence upper bound. Database-level error distribution reveals that many of the persistent errors are in databases that require deep domain-specific knowledge, such as financial, thrombosis\_prediction, and toxicology. This suggests that the remaining errors stem from a lack of understanding of specialized domain logic, which represents a direction for future work.}


\section{Conclusion}

We present DIVER, an expert-free Text-to-SQL system that automates evidence via dynamic interactive value-linking and reasoning. Using a toolbox and Chain of Thoughts and Facts (CoTF) workspace, DIVER grounds thoughts in database facts, handling ambiguous queries and complex values. Experiments show up to 10.82\% accuracy gains, with interactive linking outperforming static methods. This enhances real-world robustness and motivates future work on advanced value-linking and expert-free systems.

\bibliographystyle{unsrt}
\bibliography{custom}

\end{document}
\endinput